\documentclass[hyper,12pt,letterpaper]{JHEP3}
\usepackage{amsmath,amssymb,multirow,array}

\numberwithin{equation}{section}

\numberwithin{equation}{section}
\newcommand{\ben}{\begin{eqnarray}\displaystyle}
\newcommand{\een}{\end{eqnarray}}
\newcommand{\be}{\begin{equation}}
\newcommand{\ee}{\end{equation}}
\newcommand{\lb}{\left (}
\newcommand{\rb}{\right )}
\newcommand{\ltb}{\left [}
\newcommand{\rtb}{\right ]}

\newcommand{\nn}{\nonumber}

\newcommand{\bc}{\begin{center}}
\newcommand{\ec}{\end{center}}

\newcommand{\Rmnum}[1]{\expandafter\@slowromancap\romannumeral #1@}

\def\calm         {{\cal M}}

\def\k{\kappa}
\def\calf         {{\cal F}}
\def\tq{\tilde{q}}

\def\cg2{\cos (\pi V)}
\def\sg2{\sin (\pi V)}
\def\cb2{\cos (\delta/2)}
\def\sb2{\sin (\delta/2)}

\def\sg{r_0^2 {\rm sinh}^2\gamma }

\def\cg{r_0^2 {\rm cosh}^2\gamma }

\def\[{\left [}
\def\]{\right ]}
\def\({\left (}
\def\){\right )}

\title{The Fate of Flat Directions in Higher Derivative Gravity}
\author{Nabamita Banerjee$^{a}$\footnote{nbaner@nikhef.nl}, Suvankar
  Dutta$^b$\footnote{suvankar@iiserb.ac.in} and Ivano Lodato$^a$\footnote{ilodato@nikhef.nl}\\
  $^a$Nikhef Theory Group, Science Park 105, 1098 XG Amsterdam, The Netherlands \\
  $^b$Dept. of Physics, Indian Institute of Science Education and
  Research Bhopal, Bhopal 462 023, India}

\abstract{We discuss the fate of flat directions in higher derivative
  gravity by studying two explicit examples, namely higher derivative
  gauged supergravity in five dimensions and higher derivative type
  IIB string theory in ten dimensions. In the first case, the
  supersymmetric spinning black hole solution in asymptotically $AdS$
  spacetime, found by Gutowski and Reall, is analyzed. In this case we find
  that the flat direction at the two derivative level is not lifted
  after addition of higher derivative terms, and as it turns out, this
  result holds even for non-supersymmetric deformations of the higher
  derivative action. For the rotating D3-brane solutions in type IIB
  theory, the dilaton parametrizes a flat direction at leading order,
  but its fate changes upon including order $(\alpha')^3$
  supersymmetric higher derivative corrections to the type IIB action,
  i.e. its leading value gets fixed. }

\begin{document}
\maketitle

\section{Introduction and summary}

According to the attractor mechanism, the near horizon field
configuration of an extremal black hole is insensitive to the
asymptotic data on scalar fields of the theory. Also, many moduli
fields of the theory are fixed at the horizon, while others remain
unfixed, meaning that the black hole entropy does not depend on them.
The attractor mechanism has been observed for asymptotically flat as
well as asymptotically $AdS$ black holes and in theories with higher
derivative interactions, and there is a long list of papers where this
subject has been studied widely
\cite{9508072,9602136,9711053,9812082,9906094,0009234,
  0007195,0409148,0503219,
  0506177,0507096,0602005,0601016,0601228,0602022, 0602292,0603149,
  0604106,0605279,0606108,0711.0036}. Here, we summarize the main
points, that will be play a role in our analysis:

\begin{enumerate} 
\item Let us consider a general theory of gravity, coupled to abelian
  gauge fields, neutral scalar fields and p-form gauge fields, with a
  local Lagrangian density, which is invariant under gauge and general
  coordinate transformation. Suppose the theory admits a rotating or
  spherically symmetric extremal black hole solution. It has been
  proven \cite{0508042,0508218} that the entropy of this black hole
  remains invariant under continuous deformation of the asymptotic
  data for the moduli fields\footnote{It has been observed that there
    can be discrete jumps because of multi-centered black holes.}.

  It is important to stress that this result does not depend on the
  supersymmetries the solution preserves but relies only on the
  existence of an $AdS_2$ component in the near-horizon geometry.
  This allows us to define the ``entropy function'' \cite{0508042} of
  the theory, and by extremizing it, find the explicit values of the
  near-horizon parameters.

  In presence of higher derivative terms, even finding a generic
  solution of the full theory is a non-trivial task.  However we can
  restrict our attention on the subclass of extremal black hole
  solutions that are not destabilized by higher derivative terms,
  i.e. they still admit an $AdS_2$ component in the near-horizon
  geometry.  If that is the case, then we can expect the results for
  the attractors at the two derivative level to hold even for a
  covariant theory of higher derivative gravity.

\item {\bf Flat direction: } As we already pointed out, the attractor
  equations fix some of the moduli fields at the horizon in terms of
  the black hole charges.  On the other hand, it is possible that
  certain moduli fields cannot be fixed by extremization, meaning the
  entropy function has a series of degenerate stationary points. In
  that case, the entropy will be independent of the near-horizon
  values of these moduli, that we will refer to as flat directions.\\
  The existence of flat directions is strictly related to the
  (super)symmetries preserved by the solution, and it's likely that
  the same symmetries will completely constrain the behaviour of flat
  directions even when higher derivative terms are considered. It is
  however possible that the specific form of the higher derivative
  interactions, and, as a consequence, its symmetries, might influence
  the fate of flat directions in higher derivative gravity.

\end{enumerate}
Generically, we expect that if the two derivative theory has a BPS
black hole solution with a flat direction, then supersymmetry will
protect the structure of the near horizon geometry and the flat
direction will not be lifted, when supersymmetric higher derivative
interactions are considered.  On the other hand, nothing can be said a
priori for non-BPS solutions.  To confirm our expectations for the BPS
case and obtain some knowledge about the non-BPS case, we study two
concrete examples:

\begin{enumerate}
\item We consider five dimensional minimal gauged supergravity in presence of
  higher derivative corrections. There exists an asymptotically $AdS$,
  supersymmetric solution of two derivative gravity \cite{GutoReal1}
  that preserves 1/4 of the supersymmetries and the near-horizon
  geometry of this solution is specified in terms of a single
  parameter $\Delta$. The five dimensional field content is given by
  the metric and one gauge field. We perform a dimensional reduction
  on this five dimensional geometry over a circle ($\psi$ direction),
  obtaining a larger field content including the metric, two gauge
  fields and two scalars. We write down the entropy function and the
  attractor equations in four dimensions and find that the equations
  are satisfied without any knowledge of one scalar field and,
  moreover, the entropy is independent of its near-horizon value. This
  means that this scalar field is a flat direction of our theory.  We
  explicitly check that the flat direction remains flat, even in
  presence of higher derivative terms in the Lagrangian. The details
  are studied in section \ref{5d}.

\item Next, we condiser rotating $D3$ brane solution in type IIB string theory, which
  does not preserve any supersymmetry. The solution admits an extremal
  limit and the near-horizon geometry has an $AdS_2$ part. The
  near-horizon value of dilaton does not appear in the entropy
  function, or in the entropy, therefore it is a flat direction of the
  theory. When higher derivative terms are considered in the action,
  the fate of dilaton changes, i.e. its near-horizon value gets fixed,
  but remains independent of physical charges. The details are studied
  in section \ref{10d}.
\end{enumerate}

Finally, we end the paper by outlining possible extensions and future
projects in section \ref{futuredirections}.  We hope to report on
these questions in future. The paper also contains four appendices,
where all the technical details are provided. Appendices \ref{5d
  and 4d charges} and \ref{reduction} contain the details of five
dimensional Noether charges and Kaluza-Klein reduction
formulae. In appendix \ref{v1v2redun} we explain some issues concerning the higher derivative corrections
of the five dimensional near-horizon geometry.  In the last appendix \ref{dilatonsolution} we analyze explicitely
the solution of the dilaton equation of motion and give the value of the higher derivative invariant as a function of the near-horizon parameters.

%%%%%%%%%%%%%%%%%%%%%%%%%%%%%%%%%%%%%%%%%%%%%%%%%%%%%%%%%%%%%%%%%%%%%%%%%%%%%%%%%%%%%%%%%%%%%%%%
\section{Flat direction in five dimensional theory}
\label{5d}

Not long ago all possible purely bosonic supersymmetric solutions of
minimal gauged supergravity in $5$ dimensions were classified
\cite{GauntGuto}, using the properties of the Killing spinors. These
solutions are known to preserve $1/4$ of supersymmetries\footnote
{$AdS_5$ is the only maximally supersymmetric solution of the gauged
  theory} (only $2$ supercharges in the minimal theory) and of course,
they solve the equations of motion arising from the minimal gauged
supergravity action \cite{GuSieTown,GauntGuto}.  Analyzing all
possible near horizon geometries of these supersymmetric solutions,
Gutowski and Reall \cite{GutoReal1} were able to find an one-parameter
family of black hole solutions, which has a spatially compact horizon
(squashed $S^3$) and is (globally) asymptotically $AdS_5$, in contrast
with the ungauged case where the near horizon geometry of a BPS
solution is always maximally symmetric. We work in a suitable
coordinate system, where the $AdS_2$ part of the near horizon geometry
is manifest \cite{0902.4033}. The metric, the $U(1)$ gauge field and
its field strength have the following form:
\ben\label{nhgeoEF} ds^2 &=& v_3 \lb
B\cos\theta {d\chi } +\frac{e_0} r dr + e_0 r \ dt +{d\psi }\rb^2+v_2
\left({d\theta }^2+ \sin^2\theta \ {d\chi }^2 \right)+v_1
\left(\frac{{dr}^2}{r^2}-r^2 {dt}^2
\right),\nn\\
A&=& \left(e_0 \varphi +e_1\right) r {dt} - (B \varphi +P) \cos\theta
{d\chi }+\frac{e_0}{r} dr
+ \varphi {d\psi},\\
F&=& (e_0 \varphi+e_1) dr \wedge dt + (B \varphi+P) \sin\theta
d\theta\wedge d\chi \ .\nn 
\een

We consider, in the following, minimal gauged supergravity theory coupled to a single
$U(1)$ gauge field, including some supersymmetric higher derivative
terms. These higher derivative terms are all four derivatives and they
are related to mixed gauge gravitational Chern-Simons term by
supersymmetry. The supersymmetric completion of this term was first found
in \cite{Hanaki}, using the superconformal formalism, which gives a
complete off-shell result. An on-shell version of these higher
derivative supersymmetric invariants was derived later in
\cite{Cremonini}, by integrating out all the auxiliary fields. The
action obtained in this method, however, can be reduced further, by
means of partial integrations, field redefinitions and Bianchi
identities \cite{Myers}. Finally one can show that the action includes
only five bosonic higher derivative terms
\ben\label{act1} 
S_5 &=& \int d^5x\sqrt{-\hat{g}}\bigg[\hat{R}
+\frac{12}{L^2} -\frac{\hat{F}^2}{4} +  \frac{\kappa}{3}\epsilon ^{\mu \nu \rho
  \sigma \delta } \hat{A}_{\mu }
\hat{F}_{\nu \rho } \hat{F}_{\sigma \delta }+L^2 \bigg( c_1 \hat{R}_{\mu \nu \rho \sigma }
\hat{R}^{\mu \nu \rho \sigma
} \nn \\
&& + c_2 (\hat{F}^2)^2 + c_3 \hat{F}_{\mu }{}^{\nu } \hat{F}_{\sigma
}{}^{\mu } \hat{F}_{\nu }{}^{\rho } \hat{F}_{\rho }{}^{\sigma } + c_4
\hat{R}_{\mu \nu \rho \sigma } \hat{F}^{\mu \nu}\hat{F}^{\rho \sigma }
+ c_5 \epsilon ^{\mu \nu \rho \sigma \delta }\hat{A}_{\mu }
\hat{R}_{\nu \rho \gamma \eta } \hat{R}_{\sigma \delta }{}{}^{\gamma
  \eta }\bigg)\bigg] 
\een
where the supersymmetric values of the coefficients $c_2,c_3,c_4,c_5$
and $\kappa$ are given, in terms of $c_1$, by:
\begin{equation}\label{susyval}
 \kappa = \frac1{4\sqrt3}(1-288 c_1),\quad c_2 = \frac{c_1}{24},\quad c_3=-\frac{5c_1}{24},\quad c_4
 =-\frac{c_1}2,\quad c_5 = \frac{c_1}{2\sqrt3} .
\end{equation}

Notice that the action (\ref{act1}) is not gauge invariant, due to the presence of Chern-Simon terms. In the following sub-section
 we deal with this issue, obtaining a generalization of a result known at the two derivative level for asymptotically AdS black holes:
a relation between the 5 dimensional and the reduced, 4 dimensional, black hole charges.

\subsection{Black hole charges}
\label{chargescalc}
In order to obtain any knowledge on the behaviour of the moduli fields of the theory under consideration,
 we want to make use of Sen's entropy function formalism \cite{0506177}, which, however, is applicable 
only to gauge and diffeomorphism invariant theories, unlike (\ref{act1}). To circumvent this problem, we dimensionally reduce our five dimensional theory over a circle,
obtaining a four dimensional gauge invariant action (details about the reduction are presented in sub-section \ref{4dchargesbyef}
and appendix \ref{reduction}). Even so, the entropy of the reduced 4D black hole solution will depend on the four dimensional
charges, thus we must first find a relation linking the lower and higher dimensional charges. In this way we can still 
get the entropy of the ``original''  black hole solution as a function of the five dimensional charges.
Now, as it turns out, at two derivative level it was proven that the 5D and 4D black hole charges are exactly 
equivalent (for 5-dimensional $AdS$ solutions this result was first found in \cite{Astefanesei}). Obviously, in our 
case, we would like to find a relation between the black hole charges when higher derivative interactions are
considered. While the answer is known 
for asymptotically flat BPS black hole
\cite{Castro1,Castro2,deWitKat}, where a mismatch,
due to the gravitational Chern-Simon term, was found among
the charges, in the asymptotically $AdS$ case, no one (to the best of our knowledge)
has studied this relation. In the following three sub-sections we intend to fill this gap.
The construction is generic and in particular does not assume the
supersummetric values of the various  
coefficients mentioned in (\ref{susyval}).
\subsubsection{5 dimensional charges - Noether potential}

It is well known that for a
given local invariance of the action there exists a current $J^\mu$,
conserved on-shell:
\begin{equation}
\partial_\mu\,J^\mu (\xi,\phi)=-E_i\,\delta\phi^i
\end{equation} 
where $E_i$ indicates the equations of motion for the collection of fields
$\phi^i$, and $\xi$ is the local parameter of the invariance transformation.
Imposing on-shell conditions, the current is conserved and it can be re-written as
a total derivative of a second rank antisymmetric tensor $Q^{\mu\nu}$,
called Noether potential  
\begin{equation}
J^\mu=\partial_\nu \,Q^{\mu\nu}(\xi,\phi) .
\end{equation} 
 The integral of this
object over the horizon of black hole will give us its associated
charges.  

Let's start by calculating the current corresponding to abelian gauge
invariance of the action. The equations of motions obtained from the
Lagrangian (\ref{act1}) are as follows
\begin{align}
\label{eomA}
E^\mu_A=&\hat \nabla_\nu\Big(- \hat F^{\mu\nu} +8c_2(\hat F^2)\hat
F^{\mu\nu}-8c_3 \hat F^{\mu\rho}\hat F_{\rho\sigma}\hat F^{\sigma\nu}+
4c_4 \hat R^{\mu\nu\rho\sigma}\hat F_{\rho\sigma}\Big)\nn \\
& \qquad +\epsilon^{\mu\nu\rho\sigma\tau}\Big(\kappa \hat F_{\nu\rho}
\hat F_{\sigma\tau} +c_5R_{\nu\rho\alpha\beta} \hat
R_{\sigma\tau}{}^{\alpha\beta}\Big)
\nonumber\\
E^{\alpha\beta}_g
&=\tfrac12(\mathcal{L}-\tfrac\kappa3\epsilon^{\mu\nu\rho\sigma\tau}\hat 
A_\mu \hat F_{\nu\rho}\hat F_{\sigma\tau}
-c_5\epsilon^{\mu\nu\rho\sigma\tau}\hat A_\mu \hat
R_{\nu\rho}{}^{\gamma\delta}\hat R_{\sigma\tau\gamma\delta})\hat
g^{\alpha\beta}
\nonumber\\
&-\hat R^{\alpha\beta}+\tfrac12 \hat F^{\alpha\gamma}\hat
F^{\beta}{}_\gamma + c_1\big(-2 \hat R^{(\alpha|\nu\rho\sigma}\hat
R^{\beta)}{}_{\nu\rho\sigma}+4\hat \nabla_\rho \hat \nabla_\sigma \hat
R^{\rho(\alpha\beta)\sigma}\big)
\nonumber\\
&\quad -4c_2 \hat F^2 \hat F^{(\alpha|\gamma}\hat F^{\beta)}{}_\gamma
-4c_3\,\hat F^{(\alpha|\gamma}\hat F_{\gamma\lambda}\hat F^{\lambda\rho}\hat F_{\rho}{}^{\beta)} \nn \\
&\quad +c_4\big(3\hat R_{\nu\rho\sigma}{}^{(\alpha}\hat
F^{\beta)\sigma}\hat F^{\nu\rho} +2\hat \nabla_\rho\hat \nabla_\sigma
(\hat F^{\rho(\alpha}\hat F^{\beta)\sigma})\big)
\nonumber\\
&\quad+2c_5\varepsilon^{\mu\nu\rho\sigma(\alpha}(\hat \nabla_\lambda
\hat F_{\mu\nu}\hat R_{\rho\sigma}{}^{\lambda|\beta)} +2 \hat
F_{\mu\nu}\hat \nabla_\rho \hat R_\sigma{}^{\beta)}) .  \nonumber
\end{align}
Now, consider a symmetry variation $\delta_\xi\,A_\mu=\partial_\mu \xi$ of the Lagrangian density. Due to this variation we get the  
following  conserved gauge current $J^\mu$,
\begin{align}
J^\mu=&\sqrt{-\hat g}\Big(- \hat F^{\mu\nu}+\tfrac43 \kappa \epsilon^{\mu\nu\rho\sigma\tau}\hat A_\rho \hat F_{\sigma\tau}
+8c_2(\hat F^2)F^{\mu\nu}
+8c_3 \hat F^{\nu\rho}F_{\rho\sigma}\hat F^{\sigma\mu}\Big)\hat \partial_\nu\,\xi
+4c_4 \hat R^{\mu\nu\rho\sigma}\hat F_{\rho\sigma}
 \nn \\
& \qquad -\xi\epsilon^{\mu\nu\rho\sigma\tau}\Big(\frac{\kappa}{3} 
\hat F_{\nu\rho}\hat F_{\sigma\tau}
+c_5R_{\nu\rho\alpha\beta}\hat R_{\sigma\tau}{}^{\alpha\beta}\Big).
\end{align}
Next step is to evaluate the Noether potential $Q^{\mu\nu}$. For this,
we would need to add suitable improvement terms, to cast the current
in a form proportional to the symmetry variation
$\delta_\xi\,A_\mu=\partial_\mu \xi$: this means that $J^\mu$ would
vanish for a symmetric background, and $Q^{\mu\nu}$ would be a
conserved quantity. This procedure is rather cumbersome, but luckily,
we can obtain the same conservation law directly from the equation of
motion for the gauge field, namely: $$E^\mu_A=\nabla_\nu Q^{\mu\nu}.$$
Now, it is obvious that $Q^{\mu\nu}$ is a conserved quantity when the equation of motion for the gauge field are imposed.
Furthermore its integral over the horizon corresponds to the black hole electric charge
associated to gauge symmetry, i.e. the electric charge:
\begin{equation}\label{electriccharge}
Q_{5D}=\int d \Sigma_{\mu\nu}Q^{\mu\nu}
\end{equation}
where $d\Sigma_{\mu\nu}=dS\,\sqrt{h}\epsilon_{\mu\nu}$, $\sqrt{h}$ is
the determinant of the induced metric on the null surface S of the
horizon,and the tensor $\epsilon_{\mu\nu}$\footnote{The notation we use for the binormal and the antisymmetric epsilon tensor are similar, but they
  are tensors of different ranks.} is the binormal on that surface,
satisfying the normalization condition
$\epsilon_{\mu\nu}\epsilon^{\mu\nu}=-2$, with only one non-zero
component, $\epsilon_{tr}=v_1$.  After some manipulation, using the
Bianchi identities for the field strength and the Riemann tensor, we
obtain:
\begin{align}\label{NPE}
  Q^{\mu\nu}=&- \hat F^{\mu\nu}+2\kappa
  \epsilon^{\mu\nu\rho\sigma\tau}\hat A_\rho \hat F_{\sigma\tau}
  +8c_2(\hat F^2)F^{\mu\nu}-8c_3 \hat F^{\mu\rho}F_{\rho\sigma}\hat
  F^{\sigma\nu}
  +4c_4 \hat R^{\mu\nu\rho\sigma}F_{\rho\sigma} \nn \\
  & \qquad -4c_5\epsilon^{\mu\nu\rho\sigma\tau}\Big(\hat
  \Gamma_{\rho\beta}{}^\alpha \hat \partial_\sigma \hat
  \Gamma_{\tau\alpha}{}^\beta +\tfrac23 \hat
  \Gamma_{\rho\beta}{}^\alpha \hat \Gamma_{\sigma\gamma}{}^\beta \hat
  \Gamma_{\tau\alpha}{}^\gamma\Big) .
\end{align}
Now, using the the above expression of Noether potential in
(\ref{electriccharge}), we can compute the five dimensional conserved
electric charge for the background (\ref{nhgeoEF}). The result is quite
long: we present it in appendix \ref{5d and 4d charges}.

The evaluation of the angular momentum is completely analogous.  The
full action (\ref{act1}) is invariant under diffeomorphism and the
current associated to this invariance reads:

\begin{align}
  J^\mu(\xi,A,g)&= \Big(- \hat F^{\mu\nu}+4\frac{\kappa}{3}
  \epsilon^{\mu\nu\rho\sigma\tau}\hat A_\rho \hat F_{\sigma\tau} +4c_4
  \hat R^{\mu\nu\rho\sigma}\hat F_{\rho\sigma}
  \nonumber\\
  &\,\,+8c_2(\hat F^2)\hat F^{\mu\nu} +8c_3\hat F^{\nu\rho}\hat
  F_{\rho\sigma}\hat F^{\sigma\mu}\Big)\big(\xi^\lambda \hat
  F_{\lambda\nu}+\hat \nabla_{\!\!\nu} (\xi^\lambda \hat
  A_\lambda)\big)
  \nonumber\\
  &-2\Big(\hat g^{\big[\nu[\sigma}\hat g^{\rho]\mu\big]}
  +2\,c_5\varepsilon^{\mu\alpha\beta\sigma\tau}\hat A_\tau \hat
  R_{\alpha\beta}{}^{\nu\rho}+2\,c_1 \hat R^{\mu\nu\rho\sigma}+c_4
  \hat F^{\mu\nu}\hat F^{\rho\sigma}\Big)\hat \nabla_\rho[2\hat
  \nabla_{(\sigma}\xi_{\nu)}]
  \nonumber\\
  &+2\hat
  \nabla_\rho\Big(2\,c_5\varepsilon^{\alpha\beta\rho\sigma\tau}\hat
  A_\tau \hat R_{\alpha\beta}{}^{\mu\nu}+2\, c_1 \hat
  R^{\mu\nu\rho\sigma}+c_4\hat F^{\mu\nu}\hat
  F^{\rho\sigma}\Big)(2\hat \nabla_{(\sigma}\xi_{\nu)})
  \nonumber\\
  &-\xi^\mu\mathcal{L}'
\end{align}
where, $\xi^\mu$ is the rotational Killing vector of the black hole
spacetime.  Now, to extract a total derivative from the current, we
add a linear combination of the equations of motion of $E^\mu_A$ and
$E^{\mu\nu}_g$. As it can be easily understood, adding these term will
not alter in any way the final physical result, since they vanish
on-shell.  The correct combination we use to extract the Noether
potential for diffeomorphism is:

\begin{equation}
 J^\mu+2E^{\mu\nu}_g\xi_\nu+(\xi\cdot A) E^\mu_A=\nabla_\nu \Theta^{\mu\nu} 
\end{equation}
where
\begin{align}\label{NPAM}
  \Theta^{\mu\nu}&=\big(-\hat F^{\mu\nu} +4c_4 \hat
  R^{\mu\nu\rho\sigma}\hat F_{\rho\sigma}+8c_2(\hat F^2)\hat
  F^{\mu\nu} -8c_3\hat F^{\mu\rho}\hat F_{\rho\sigma}\hat
  F^{\sigma\nu} +4\frac{\kappa}{3} \epsilon^{\mu\nu\rho\sigma\tau}\hat
  A_\rho \hat F_{\sigma\tau}\big)(\xi\cdot A)
  \nonumber\\
  &\qquad+c_5\Big(4\epsilon^{\mu\nu\rho\alpha\beta}\hat A_\rho \hat
  R_{\alpha\beta}{}^{\sigma\tau}\hat \nabla_\tau\xi_\sigma
  +2\epsilon^{\mu\rho\sigma\alpha\beta}\hat F_{\rho\sigma} \hat
  R_{\alpha\beta}{}^{\nu\tau}\xi_\tau
  +4\epsilon^{\rho\sigma\alpha\beta(\nu}\hat F_{\rho\sigma}\hat
  R_{\alpha\beta}{}^{\tau)\mu}\xi_\tau\Big)
  \nonumber\\
  &\qquad-2\hat g^{\big[\nu[\sigma}\hat g^{\rho]\mu\big]}\hat
  \nabla_\rho\xi_\sigma +c_1\Big(-4\hat R^{\mu\nu\rho\sigma}\hat
  \nabla_\rho\xi_\sigma +8\hat \nabla_\rho(\hat
  R^{\mu\nu\rho\sigma})\xi_\sigma\Big)
  \nonumber\\
  &\qquad+c_4\big(-2\hat F^{\mu\nu}\hat F^{\rho\sigma}\hat
  \nabla_\rho\xi_\sigma +2\hat \nabla_\rho(\hat F^{\mu\rho}\hat
  F^{\nu\sigma})\xi_\sigma +4\hat \nabla_\rho(\hat F^{\mu(\sigma}\hat
  F^{\rho|\nu)})\xi_{\sigma}\big) . 
\end{align}
With the above expression of the Noether potential, the conserved
angular momentum can be computed using:
\begin{equation}\label{angularmomenta}
\Theta_{5D}=\int d \Sigma_{\mu\nu}\Theta^{\mu\nu}.
\end{equation}
Again, the result is presented in appendix \ref{5d and 4d charges}.

\subsubsection{4 dimensional charges - Dimensional reduction and
  entropy function}\label{4dchargesbyef} 

In this subsection, we determine the four dimensional charges for the
corresponding four dimensional system using the entropy function formalism. 
The derivation is, again, completely generic as it only depends on the form
of the near horizon solution presented in (\ref{nhgeoEF}) and does not
assume any particular value for any near horizon parameters.   

As explained above, to apply entropy function, we need gauge invariant
Lagrangian and thus, we first need to perform a Kaluza-Klein reduction
of the action (\ref{act1}). We take the following ansatz for the
metric and gauge field for the reduction,
\ben\label{kkans}
ds^2 &=& \hat{g}_{\mu\nu}dx^{\mu}dx^{\nu} = h_{ij}dx^idx^j+\phi(r)
(d\psi+ A_{kk})^2, \nn\\
\hat{A}_{\mu} &=& A_idx^i +\sigma(r) (d\psi+ A_{kk})
\een
and compactify along $\psi$ direction. We denote $A_{kk}$ as the
Kaluza-Klein (KK) gauge field and the corresponding field strength is
given by
$(F_{kk})_{ij}=\partial_i(A_{kk})_j-\partial_j(A_{kk})_i$. The four
dimensional gauge field is $A$ and the corresponding field strength is
$F$.  All un-hatted curvature quantities are composed of the four
dimensional metric $h_{ij}$.  The reduction of gauge invariant terms
in the action is straightforward (appendix \ref{reduction}).  On the
other hand, the reduction of the two Chern-Simons terms, which are
gauge non-invariant, is tricky, and requires the addition of some total derivatives terms.
 Here, we present only the reduced four dimensional action
corresponding to these two gauge non-invariant terms.  Specifically
the two derivative gauge Chern-Simon term takes the following form
\ben\label{GCS} 
\frac{\kappa}{3}\int
\sqrt{-g}\epsilon^{\mu\nu\rho\sigma\delta} \hat A_{\mu} \hat
F_{\nu\rho} \hat F_{\sigma \delta} &=& 4 \pi \kappa \int d^4x
\sqrt{-h} \epsilon^{abcd} \bigg[\frac{1}{3} \sigma^3
(F_{kk})_{ab}(F_{kk})_{cd}+ \sigma^2 (F_{kk})_{ab}F_{cd}
+ \sigma F_{ab}F_{cd}\bigg], \nn \\
\een
while the four derivative mixed Chern-Simon term reads:
\be
c_5 \int \sqrt{-g}\epsilon^{\mu \nu \rho \sigma \delta} \hat A_{\mu}
\hat R_{\nu\rho}\,^{\alpha \beta}  
\hat R_{\sigma \delta \alpha \beta} = c_5(T_1- 4 T_2) ,
\ee
where, $T_1$ and $T_2$ are given by,
\ben
T_1 &=& 4 \pi \int d^4x \sqrt{-h} \epsilon^{ijkl}\bigg[\sigma \bigg( R^{mn}\,_{ij} 
\bigg(R_{klmn}- \Phi \big((F_{kk})_{km}(F_{kk})_{ln} + (F_{kk})_{kl}(F_{kk})_{mn} \big)\bigg) \nn\\
&&+\frac{\Phi^2}{4}\bigg((F_{kk})_{ij}(F_{kk})_{kl}(F_{kk})^2-
2 (F_{kk})_{ij}(F_{kk})_{km}(F_{kk})^{mn}(F_{kk})_{nl}\bigg) \nn\\
&&+\frac{\Phi}{2}\bigg(\nabla_m (F_{kk})_{ij}\nabla^m (F_{kk})_{kl}\bigg)
\bigg) \bigg] \nn\\
T_2 &=& 2 \pi  \int d^4x \sqrt{-h} \epsilon^{ijkl}F_{kl}\bigg[ \frac{\Phi}{2} R^{mn} \, _{ij} (F_{kk})_{mn}
+ \frac{\Phi^2}{4}(F_{kk})_{im}(F_{kk})_{jn}(F_{kk})^{mn}-\frac{\Phi^2}{8}(F_{kk})_{ij}(F_{kk})^2\bigg]\nn
\een
and $(F_{kk})^2= (F_{kk})_{ab}(F_{kk})^{ab}$. The periodicity of the compact 
direction is $4 \pi$.\\
 Now, the four
dimensional action obtained is gauge invariant so that we can apply the entropy function
formalism. Our ansatz for near-horizon metric and 
gauge field is (\ref{nhgeoEF}).\\
%We define the entropy function in the following way.  
From the four dimensional point of view we have two gauge fields, one coming from
usual five dimensional gauge field and the other coming from
the metric components $g_{\psi\mu}$ (Kaluza-Klein gauge field): to each of those gauge fields 
corresponds a charge, respectively $Q$ and $\Theta$, to which we associate a charge parameter, $e_1$ and $e_0$ respectively.
\\
The entropy function is, then, defined as follows: 
\be {\cal E} = 2\pi (Q
e_1 + \Theta e_0 - \bar L) \ee where, $\bar L$ is given by, \be \bar L
= \frac{8 \pi^2}{16 \pi G_5}\int d\theta d\chi {\cal L}_4 
\ee 
and ${\cal L}_4$ is the reduced four dimensional Lagrangian,including
the higher derivative terms.

The attractor equations are obtained by minimizing the entropy
function with respect to the near-horizon parameters, while the four
dimensional physical charges are given by:
\ben
Q= \frac{\partial {\bar L}}{\partial {\bar e_1}},\qquad
\Theta = \frac{\partial {\bar L}}{\partial {\bar e_0}}.
\een
Calculating ${\bar L}$ over the near-horizon geometry (\ref{nhgeoEF}),
we find the expression for the four dimensional physical charges.  The
proper four dimensional charges are rescaled as in \cite{0506177} to
$\tilde Q= 2 Q, \ \tilde \Theta= 2 \Theta$.

\subsubsection{Relation Between 5D-4D Charge}

Comparing the five dimensional charges with the corresponding four
dimensional charges, we find a
complete match between the two at two derivative level. However, as already seen
for asymptotically flat black hole solutions, they
differ at the four derivative level and the difference is proportional
to $c_5$ only (gravitational Chern-Simon term). We also find that the
relation between five-dimensional and four-dimensional charge remains
exactly the same as in the case of asymptotically flat black holes. For
the angular momentum $\Theta_{5D}$, our result is more generic than
the one in \cite{Banerjee}, as in our case the black hole can carry an
extra parameter $P$.\footnote{$P$ can be thought of as a magnetic
  field.} Thus, we see that the difference between five and four
dimensional charges is purely a topological effect due to the
Chern-Simon term which is not gauge invariant. The details can be
found in \cite{Banerjee}. The bottom line of our analysis is that the
asymptotic geometry of the space time is not relevant to account for
the differences in the physical charges. The relations between 4D and 5D
charges for asymptotically $AdS$ black holes reads:
%\footnote{The only known
%  example of such black hole is the spinning black hole of
%  Gutowski-Reall and they carry non trivial magnetic field.} are given
\ben\label{5d4drelation} \tilde Q = - Q_{5D} +\frac{8 \pi B v_3}{
  G v_2} c_5 , \quad \tilde \Theta= - \Theta_{5D} + \frac{4 \pi P
  v_3^2(e_0^2 v_2^2-B^2 v_1^2)}{G v_1^2 v_2^2} c_5 \een

These expressions are one of the main results of our paper and constitute a generalization 
of previous work \cite{Astefanesei} to higher-derivative gravity theory.

\subsection{Flat directions in higher derivative gravity}

So far we have studied the generic relation between four dimensional
and five dimensional charges. Now we concentrate on a particular class
of supersymmetric solutions in five dimensions and find that it
exhibits a flat direction. Our goal is to study the fate of this flat
direction when higher derivative interactions are taken into account.

We consider, in the following, the supersymmetric asymptotically $AdS_5$ black hole solution
presented by Gutowski and Reall \cite{GutoReal1}. This solution is
$1/4$ BPS (preserves $2$ supercharges) and its near-horizon geometry
has an $AdS_2$ component. We use the coordinates that make the $AdS_2$
part of the near horizon geometry manifest \cite{0902.4033},
\ben \label{nhgeoGR}
ds^2 &=& \frac{1}{\Delta ^2+9L^{-2}} \left(\frac{{dr}^2}{r^2}-{dt}^2
  r^2\right)+\frac{1}{\Delta ^2-3L^{-2}}\left( {d\theta }^2+{d\chi}^2
  \sin ^2(\theta ) \right)\nn\\
&&+\left(\frac{\Delta}{\Delta
    ^2-3L^{-2}}\right)^2 \left(d\psi+\cos\theta d\chi -\frac{3 r }{L
    \Delta} \frac{\Delta^2-3L^{-2}}{\Delta ^2+9L^{-2}}\left(dt +
    \frac{dr}{r^2}\right)\right)^2\nn\\
%A&=& \frac{\sqrt{3} \Delta}{\Delta ^2+9L^{-2}} r \left(
%  dt-\frac{dr}{r^2} \right)
%-\frac{\sqrt{3}\cos\theta}{L(\Delta^2-3L^{-2})} d\chi
F&=& \frac{\sqrt{3} \Delta}{\Delta ^2+9L^{-2}} dr\wedge dt  +
\frac{\sqrt{3}\sin\theta}{L(\Delta^2-3L^{-2})} d\theta\wedge d\chi
\een
Comparing this solution with near-horizon ansatz given in
(\ref{nhgeoEF}) we find the following values of the near-horizon
parameters 
\ben\label{leadingsolutions5d} v_1 &=& \frac{1}{\Delta ^2+9L^{-2}},
\qquad v_2 = \frac{1}{\Delta ^2-3L^{-2}}\ , \qquad v_3 = \lb
\frac{\Delta}{\Delta^2-3L^{-2}}\rb^2\ , \nn\\
e_0 &=& -\frac{3 }{L \Delta} \frac{\Delta^2-3L^{-2}}{\Delta
  ^2+9L^{-2}}, \qquad B=1,
\qquad e_5 = e_1 +e_0 \varphi=   \frac{\sqrt{3} \Delta}{\Delta ^2+9L^{-2}},\nn \\
&& \qquad \qquad A_{\chi} = P+ B \varphi =
\frac{\sqrt{3}}{L(\Delta^2-3L^{-2})}.  
\een

%As mentioned in the previous subsection, the presence of gauge
%non-invariant terms in the action forces us to reduce the theory over
%a circle ($\psi$ direction) to make it gauge invariant. Then only we
%can apply entropy function formalism in four dimensions. However from
%four dimensional point of view we have a metric, two gauge fields and
%two scalars $v_3$ and $\varphi$.  We write down the entropy function
%and attractor equations and solve them to find the near-horizon
%geometry.

The leading attractor equations (i.e. derived from the two derivative
action) are given by, 
\ben \label{leadingnhgeometry5d} v_1 \
\text{equation :} && v_1^2 \left(B^2 \left(\varphi ^2+v_3^2\right)-2
  \Lambda
  v_2^2-4 v_2\right)+2 B P \varphi  v_1^2\nn\\
&& +v_2^2 \left(e_0^2 \left(\varphi
    ^2+v_3^2\right)+2 e_1 e_0 \varphi +e_1^2\right)+P^2 v_1^2=0 \nn\\
%\een
%\ben
\nn\\
v_2\ \text{equation :}  && B^2 \varphi ^2 v_1^2+B^2 v_3^2 v_1^2+2 B P \varphi 
   v_1^2+e_0^2 \varphi ^2 v_2^2 -4 v_2^2 v_1\nn\\
&& +2 e_0 e_1 \varphi  v_2^2+e_1^2 v_2^2+e_0^2 v_2^2
   v_3^2+P^2 v_1^2+2 \Lambda  v_2^2 v_1^2= 0\nn\\
%\een
%\ben
\nn\\
v_3 \ \text{equation :}  && v_1^2 \left(B^2 \left(\varphi ^2+3
    v_3^2\right)-2 \Lambda  v_2^2-4 v_2\right)+2 B P 
   \varphi  v_1^2\nn\\
&& -v_2^2 \left(e_0^2 \left(\varphi ^2+3 v_3^2\right)+2 e_1 e_0 \varphi
   +e_1^2\right)+P^2 v_1^2+4 v_2^2 v_1=0\nn\\
%\een
%\ben
\nn\\
\varphi \ \text{equation :} && B^2 \varphi  v_3 v_1^2+B v_1 \left(24
  \varphi  v_2 \left(e_0 \varphi +e_1\right) \kappa+P v_1 v_3\right)\nn\\
&&-v_2 \left(e_0 \varphi +e_1\right) \left(e_0 v_2 v_3-24 P v_1
   \kappa\right) = 0
\een
Other two attractor equations (for $e_0$ and $e_1$) define the four
dimensional charges $\Theta_{4D}$ and $Q_{4D}$ in terms of
near-horizon geometry.

Substituting the leading values of near-horizon geometry
(\ref{leadingnhgeometry5d}) in the attractor equations one can check that
the first three equations ( corresponding to $v_1, v_2$ and $v_3$)
vanish. Furthermore the $\varphi$ equation vanishes for a particular
value of $\kappa= \frac{1}{4\sqrt{3}}$, which is the supersymmetric
value. Thus one does not need any specific
$\varphi$ to solve the attractor equations. It is also easy to check that
the entropy of the black hole does not depend on the near-horizon
value of this scalar field and it is given by: 
\be 
S= \frac{2 \pi
  \Delta L^4}{G \left(\Delta ^2 L^2-3\right)^2}.  
\ee 

Therefore we conclude that, at two derivative level, $\varphi$ is a flat direction, 
as it was already observed in \cite{Astefanesei}.

%We want to check the fate of this flat direction when we add higher
%derivative terms to the Lagrangian. 

\subsubsection{Higher derivative correction to the extremal
  near-horizon geometry and fate of the flat direction} 

To check whether $\varphi$ remains flat in presence of higher
derivative terms we follow the same procedure as we did at leading
order. We first find the corrections to the five
dimensional near-horizon geometry due to higher derivative interactions. Then we study the attractor
equations derived from the entropy function in presence of higher
derivative terms on this corrected solution.

We consider the following higher derivative correction to the five
dimensional near-horizon geometry\footnote{We do not consider any correction
for $v_3, A_{\chi}$ and $B$ as one can use redundancies in the leading
solution to choose the corrections to the above three parametersto be
zero. We address this issue in appendix \ref{v1v2redun}.}:
\ben\label{corrections5d}
v_1 &=& \frac{1}{\Delta ^2+9L^{-2}} +\gamma V_1, \qquad v_2 =
\frac{1}{\Delta ^2-3L^{-2}}+\gamma V_2,\qquad
v_3 = \frac{\Delta}{\Delta^2-3L^{-2}} ,\nn\\ 
e_0 &=& -\frac{3 }{L
  \Delta} \frac{\Delta^2-3L^{-2}}{\Delta ^2+9L^{-2}} +\gamma E_0,
\qquad 
e_5 = \frac{\sqrt{3} \Delta}{\Delta ^2+9L^{-2}}+\gamma E_5,\nn\\
A_{\chi}&=& \frac{\sqrt{3}}{L(\Delta^2-3L^{-2})}, \qquad
B= 1 .
\een

We can solve for these higher derivative corrections ($V_1, V_2, E_0$
and $E_5$) using the five dimensional equations of motion. The
solution is given in appendix \ref{v1v2redun}. One can easily check that
the fifth component of five dimensional gauge field ($\varphi$) never
appeared in any Einstein's equation. Therefore, we can not fix this
scalar or its higher derivative correction in five dimensions. 
% Thus, for a five dimensional theory,
%this field remains unfixed. 
However, it is important to check whether the
entropy depends on this scalar field or not. To this end, we first
verify that the corrected five dimensional solution solves the higher
derivative attractor equations and then compute the entropy function
on this solution. As it turns out, imposing on-shell conditions on the entropy function 
will make its dependence on $\varphi$ disappear. In fact, the entropy reads:

\ben
S= \frac{2 \pi  \Delta  L^4}{G \left(\Delta ^2 L^2-3\right)^2}\bigg[1
&+&\frac{ \gamma}{ \left(\Delta ^4 L^4-6 \Delta ^2 L^2-15\right)}
\bigg\{72 \bigg(2 c_2 (7 \Delta ^4 L^4-10 \Delta ^2
   L^2 +3) \nonumber\\
&+&c_3 (9 \Delta ^4 L^4+2 \Delta ^2 L^2+3)+2 \sqrt{3} c_5
   (\Delta ^4 L^4-7 \Delta ^2 L^2-5)\bigg)\nonumber\\
&+&c_1 (-16 \Delta ^6 L^6+503
   \Delta ^4 L^4 +246 \Delta ^2 L^2+585)-24 c_4 (\Delta ^6 L^6\nn\\
&-&24
   \Delta ^4   L^4-53 \Delta ^2 L^2+15)\bigg\}\bigg] + {\cal O }(\gamma^2) 
\een

As expected, the flat direction at the two derivative level is not lifted, if the solution preserves some supersymmetries. 
What strikes as a surprise, however, is that throughout the whole analysis
 we never specified the supersymmetric values for the coefficients $c_i{}'s$ of the higher derivative interactions in 
(\ref{susyval}), meaning that the flat direction will remain flat even for  non supersymmetric deformations of the
higher derivative action (\ref{act1}). Unexpectedly, the symmetries of the leading order black hole solution seem to protect the flat directions from being lifted
independently of the symmetries of the full action.

For supersymmetric values given in (\ref{susyval}) the entropy has the
following form (remember that the $AdS$ radius $L$ also picks up a
correction):
\ben S_{susy} &=& \frac{2 \pi \Delta K L^4}{G \left(\Delta ^2
    L^2-3\right)^2}-\frac{\pi \gamma c_1 \Delta K L^4 \left(2 \Delta
    ^8 L^8-103 \Delta ^6 L^6-2633 \Delta ^4 L^4+8463 \Delta ^2
    L^2-33\right)}{G \left(\Delta ^2 L^2-3\right)^3 \left(\Delta ^4
    L^4-6 \Delta ^2 L^2-15\right)}\nn
 \een

\section{Flat direction in ten dimensional theory}\label{10d}

Supersymmetric, asymptotically $AdS_5$ black hole solution, like the
one analyzed in the previous section, has been certainly used for a
huge number of applications in the $AdS/CFT$ correspondence. However,
these are not the only solutions used to obtain some
knowledge of the dual field theory.\\
 The first attempt of
finding BPS black holes in 5 dimensional minimal gauged supergravity
dates back few years \cite{Behrndt2}, but all the solutions suffer
from not having regular horizons or naked singularity.  Later, they
were found as a special limit of a more general class of non supersymmetric
black hole solutions \cite{Behrndt}, which contain a non-extremality parameter
$\mu$ linking solutions of the ungauged theory with supersymmetric
solution of the gauged theory ($\mu=0$ is the BPS-saturated
limit). Such non supersymmetric black hole solutions of the minimal five
dimensional gauged $U(1)^3$ supergravity, which are asymptotically $AdS_5\times S^5$,
 can have a regular extreme
limit with zero Hawking temperature and finite
entropy\cite{Behrndt}. It is also possible to embed such solutions in
type IIB supergravity \cite{Cvetic}.
 The full
class of solutions\footnote{See also \cite{Buchel}.}, which was shown
to satisfy the 10 dimensional equations of motion coming from the two
derivative type IIB supergravity action, includes the 10 dimensional black hole metric, 
a self-dual five form $F_5$, three gauge fields
$a_i$, coming from the 5 dimensional $U(1)^3$ gauged theory lifted in
10 dimensions and physical charges $\tilde{q}_i$:
\begin{equation}
\label{metunc}
\begin{split}
ds_{10}^2=&\sqrt{\triangle}\left[-(H_1H_2H_3)^{-1}fdt^2
+\left(f^{-1}dr^2+r^2(d\calm_3)^2\right)\right]\\
&+\frac{1}{\sqrt{\triangle}}\sum_{i=1}^3L^2 H_i\left(d\mu_i^2
+\mu_i^2\left[d\phi_i+a_i\ dt\right]^2\right)\,,
\end{split}
\end{equation}
where $\calm_3=\{R^3,S^3\}$ is a spatial manifold corresponding to
curvatures $\k=\{0,1\}$,
\begin{equation}
\begin{split}
&a_i=\frac{\tilde{q}_i}{q_i}\ L^{-1}\left(H_i^{-1}-1\right)\,,
\qquad H_i=1+\frac{q_i}{r^2}\,, \\
&\triangle=H_1H_2H_3\sum_{i=1}^3\frac{\mu_i^2}{H_i}\,,
\qquad f=\k-\frac{\mu}{r^2}+\frac{r^2}{L^2}H_1H_2H_3\,,
\end{split}
\end{equation}
and
\begin{equation}\label{defmu}
\mu_1=\cos\theta_1\,,\qquad \mu_2=\sin\theta_1\ \cos\theta_2\,,\qquad
\mu_3=\sin\theta_1\ \sin\theta_2\, .
\end{equation}
$\kappa=0$ corresponds to flat horizon. For $\kappa=1$ horizon
topology is $S^3$ and $\kappa=-1$ gives negatively curved horizon.
The physical charges $\tq_i$ are related to charge parameters $q_i$ in
the following way,
\begin{equation}
\tilde{q}_i=\sqrt{q_i(\mu+\k q_i)}\,.
\end{equation}
 The five form field strength is given by:
\begin{equation}\label{5form}
F_5=\calf_5+\star \calf_5\,,\qquad \calf_5=d B_4\,,
\end{equation}
where,
\begin{equation}
\label{B4}
B_4=-\frac{r^4}{L}\triangle\ dt\wedge dvol_{\calm_3}-L\sum_{i=1}^3
\tilde{q}_i\mu_i^2\left(Ld\phi_i-\frac{q_i}{\tilde{q}_i}\ dt\right)
\wedge dvol_{\calm_3}\,,
\end{equation}
%dx\wedge dy\wedge dz
where $dvol_{\calm_3}$ is a volume form on $\calm_3$.  Note that the
10 dimensional Bianchi identity on the five form $\nabla_a
F^{abcde}=0$ gives rise to the five dimensional equations of motion
for the scalars and the gauge fields. Finally, the dilaton equation of 
motion admit a general solution of the form $\phi(r)=c_0+c_1 h(r)$ where
 the function $h(r)$ is singular at the horizon.
To circumvent this problem we can set $c_1=0$, so that
 the dilaton is just an arbitrary constant.\\Once again, we remind the
 readers that the above solution does not preserve any
supersymmetry. Nevertheless we will focus our analysis on it, since we
are not aware of any asymptotically $AdS$ black hole solution in 10-dimensions
that preserves some supersymmetry.

It is important to stress that the dilaton is constant and it is
not possible to find its value by solving Einstein's equations of
motion. The entropy of this black hole solution does not depend on
it. Therefore the dilation is a flat direction at the two derivatives
level. We would like to see the fate of this flat direction when we
add higher derivative terms in the action. However, for our purposes
it's easier to consider, without loss of generality, the extremal
limit of this black hole solution and, once again, consider only the
near-horizon geometry. This procedure is discussed in the following
section.

\subsection{Extremal near-horizon geometry}

The extremal limit corresponds to (taking Hawking temperature to
zero), 
\be 2r_0^6 + r_0^4(\kappa+q_1 +q_2+q_3)-q_1 q_2 q_3=0 
\ee
where $r_0$ solves the above equation given the charges.  
\\
The mass
parameter $\mu$, in the near-horizon geometry is fixed to be:
\be \mu= \k r_0^2+{r_0^4 \over L^2}\prod(1+{q_i\over r_0^2}).  
\ee

\subsubsection{Three equal charges}

For simplicity we consider only three equal charge solution:
$q_1=q_2=q_3=q$. In that case we see,
\ben
H_1=H_2=H_3&=&H=1+{q\over r^2}\\ \nn
\triangle &=& H^2,
\een
which leads to $q=2\,r_0^2$ and $\mu=27\,r_0^4$.
For convenience we take $L=1$ throughout this section.
Therefore the three equal charge black hole metric becomes,
\ben
ds_{10}^2&=&\sqrt{\triangle}\left[-(H)^{-3}fdt^2+\left(f^{-1}dr^2
    +r^2(d\calm_3)^2\right)\right]
\nonumber\\
&& + \left( d\theta_1^2 + \sin^2\theta_1 d\theta_2^2 \right)
+\sum_{i=1}^3\mu_i^2\left[d\phi_i+a_i\
  dt\right]^2 .
\een
We consider $\kappa=0$ case, i.e. flat horizon.  The analysis can be
repeated for the $\kappa=1$ case, and it's completely analogous.

The extremal near-horizon metric is given by, 
\ben\label{extremalnh1}
ds^2 &=& \frac{1}{12}\lb -r^2 dt^2+\frac{dr^2}{r^2} \rb + 3 r_0^2
(dx^2+dy^2+dz^2) + \lb d\theta_1^2 + \sin^2\theta_1 d\theta_2^2\rb
\nn\\
&& + \mu_1^2 \lb d\phi_1 + \frac{r}{3\sqrt2} dt\rb^2+\mu_2^2 \lb
d\phi_2 + \frac{r}{3\sqrt2} dt\rb^2 + \mu_3^2 \lb d\phi_3 +
\frac{r}{3\sqrt2} dt\rb^2, 
\een 
and the four form field $B$ reads,
\be\label{extremalnh2}
 B_4= -\sqrt3 r_0^3 \lb r dt + \cos^2\theta_1
d\phi_1 +\sin^2\theta_2 (d\phi_2 \cos^2\theta_2 + d\phi_3
\sin^2\theta_2) \rb \wedge dx\wedge dy \wedge dz\ .  
\ee 
Thus we see that at the extremal limit the near-horizon geometry
admits an $AdS_2$ part. However, we shall determine this near-horizon
geometry using again the entropy function analysis and discuss the fate of the
flat direction $\phi$.

\subsection{The entropy function}

We consider $x$, $y$ and $z$ direction to be compactified on a three
torus.  Therefore from the seven dimensional point of view the four
form R-R field $B_4$ appears to be an one form field $A_{\mu} =
(B_4)_{\mu xyz}$, where $\mu$ runs over all indices except $x, y$ and
$z$ (the $D3$ brane is a point-like object and $A_1$ is electrically
coupled to it).

Now we would like to compactify over $\phi_i$ directions. Therefore
from the four dimensional ($\{t,r,\theta_1,\theta_2\}$) point of view
there are three KK gauge fields $a_i=z_1 r dt$ (all of them are equal
for the three equal charges case) and six scalars: three of them coming
from the metric and three of them from $B_4$. Given the symmetry of
the problem, the scalars coming from the metric are equal and will be
denoted by $w_1$. Analogously, the scalars coming from the $4$-form
can be all denoted by $b$. In fact even starting with different values
for the scalars, they will be constrained to be equal. Therefore we
can write down the following near-horizon ansatz for the metric and
the gauge field\footnote{In fact the derivation is more involved. Once
  we break the $SO(6)$ symmetry, the lower dimensional scalars and
  gauge fields depend on the angular direction of lower dimensional
  spacetime, in this case on $\theta_1$ and $\theta_2$. One has to
  solve the attractor equations to find the angular dependence of the
  lower dimensional fields. See \cite{Sen} for the details. However in
  this case, as the leading near-horizon geometry is known, we
  substitute the angular dependence of the fields from the
  beginning. Therefore the scalars and the different components of the gauge
  fields are determined by $AdS_2$ symmetry only.}:
\ben\label{nhmet}
ds^2 &=& v_1 \lb -\rho^2 dt^2 + {d\rho^2 \over \rho^2} \rb + v_2\
(dx^2 +dy^2 +dz^2)\\ \nn 
&&+ w_1\ltb \left( d\theta_1^2 + \sin^2\theta_1 d\theta_2^2 \right)
+\sum_{i=1}^3 \mu_i^2\left[d\phi_i +z_1 \ r \ dt\right]^2\rtb 
\een
We can decompose the
seven dimensional field $A_1$ in terms of four dimensional field
$A^{(4)}$ and KK fields as follows
\ben
A = \frac{b}{2}\ \sum_i \mu_i^2 (d\phi_i + z_1 r dt) + A^{(4)}
\een
with
\be
A^{(4)}= e_0 r dt \wedge
 dz\ .
\ee
Explicitely, $dA_1$, which can be thought of as a field strength in
four dimensions, reads: 
\ben
dA &=& \bigg[q_5 \ \text{dr}\wedge \text{dt}  + b\   \sin \left(\theta _1\right) \cos
   \left(\theta _1\right) \left(-\text{d$\phi $}_3 \sin
   ^2\left(\theta _2\right)-\text{d$\phi $}_2 \cos
   ^2\left(\theta _2\right)+\text{d$\phi $}_1\right)\wedge \text{d$\theta $}_1 \nn \\
&& -b\
   \text{d$\theta $}_2 \wedge \left(\text{d$\phi $}_2-\text{d$\phi
   $}_3\right) \sin \left(\theta _2\right) \sin ^2\left(\theta
   _1\right) \cos \left(\theta _2\right)\bigg]\ ,
\een
where
\be
q_5 = e_0 + \frac{z_1 b}{2}\ .
\ee
In ten dimensions the five form RR field strength $F_5$ is given by,
\be
F_5 = dB_4 + \star dB_4
\ee
where $dB_4 = dA\wedge dx\wedge dy \wedge dz$. Therefore, the
corresponding $F_5^2$ equals:
\be
{1\over 4 \ 5!} F_5^2 = \frac{1}{2v_2^3}\lb-{q_5^2 \over v_1^2} + {2
  b^2 \over w_1^2}\rb\ .
\ee
Hence, the final result for the on-shell action reads,
\ben S&=& \frac{V_3}{16 \pi G_{10}} \int_{0}^{\pi/2}
d\theta_1\int_{0}^{\pi/2} d \theta_2 \sqrt{-g_{10}} \ltb R_{10} -
{\frac 12} (\partial \phi)^2 - \frac{1}{4 \cdot 5!}  (F_5)^2
\rtb\nn \\
&=& \frac{V_3}{128 \pi G_{10}}\frac{ v_1 v_2^{\frac32}
  w_1^{\frac52}}{2}\left(-\frac{\frac{2
      b^2}{w_1^2}-\frac{\left(\frac{b
          z_1}{2}+e_0\right){}^2}{v_1^2}}{v_2^3}+\frac{w_1 z_1^2-4
    v_1}{v_1^2}+\frac{40}{w_1}\right) 
\een

We define the entropy function with respect to the lower dimensional
charges,
\ben
{\cal E} &=& 2\pi {\cal A} \lb Q\ e_0 + \Theta \ z_1 - {S\over {\cal A}} \rb,
\een
where ${\cal A} =  \frac{V_3}{128 \pi G_{10}}$.
Solving the attractor equations we find the following solution for the
near horizon geometry:
\ben 
v_1 &=& {\sqrt{-Q}\over 24}, \quad v_2 = {(2 \Theta/3)^{2/3}
  \over (-Q)^{5/6}}, \quad z_1 = {1\over 3 \sqrt{2}} \nn\\
&& b = \frac{4\Theta}{3 Q}, \quad w_1 = \frac{\sqrt{-Q}}{2}.  \een
Furthermore, the entropy is given by:
\be
{\cal S} =\frac{\Theta}{192 \sqrt2 G_{10}}\ .
\ee

Since we know the extremal near-horizon geometry exactly
(\ref{extremalnh1}, \ref{extremalnh2}), we can solve the attractor
equations and find the lower dimensional charges in terms of a single
parameter $r_0$. We can then re-write the entropy as a function of $r_0$
and check it is in agreement with Bekenstein-Hawking law.\\
 The near-horizon geometry reads:
\ben
\label{nhparam}
v_1 &=& {1\over 12},\quad  v_2 = 3r_0^2, \quad z_1 = {1\over 3 \sqrt{2}}\nn\\
 && b= - 6\sqrt{6} \ r_0^3, \quad w_1 = 1
\een
Substituting these values in the attractor equations we get,
\be
Q=-4, \quad \Theta = 18 \sqrt6 r_0^3\ .
\ee
Therefore the entropy turns out to be
\be
{\cal S}= \frac{3\sqrt{3} r_0^3} {32 G_{10}} =
\frac{\text{Area}}{4G_{10}}\ .
\ee

One should note that the near-horizon value of the dilaton does not
appear in the entropy function, therefore it is a flat direction. Our
goal is to check what happens to this flat direction when we add
supersymmetric higher derivative terms which appear in type IIB string
theory.

\subsection{Higher derivative terms in type IIB string theory}

For type IIB supergravity, which is a low-momentum expansion of type
IIB superstring theory, the higher derivative corrections can be
written as a series in $\alpha^\prime$.  The series is of the
following form:
\begin{equation}
  \alpha^{\prime\,4}\,S_{IIB}=S^{(0)}+\alpha^\prime\,S^{(1)}+\cdots
  +(\alpha^\prime)^n\,S^{(n)},
\end{equation} 
where the first non-zero term, expected to appear at tree-level or
1-loop in the string coupling, is of the order
$\alpha^{\prime\,3}$. It is an eight derivative action, one of the
terms being the well known $R^4$ term.  Unfortunately, the standard
superfield techniques (\cite{deHaro,Rajaraman}) can not be used
for the construction of the full $S^{(3)}$ contribution to the two
derivative action, that corresponds to the supersymmetric completion
of the $R^4$ term \cite{Green2}.  Nevertheless, if one considers only
a subset of the full field content of type IIB theory, specifically
the metric and the five-form, then a general formula for the
supersymmetric higher derivative correction exists \cite{Rajaraman}.
For the sake of completeness, we outline the steps taken to obtain
such an invariant.  First of all, $U(1)$ gauge invariance of the
theory allows us to separate all the higher derivative terms by their
charge. We will then only look for terms that are neutral under $U(1)$
and contain at most one fermion bilinear.  These terms, schematically,
look like:
\begin{equation}
\label{B}
S^{(3)}_{0;B}=\int d^{10}x\,f^{(0,0)}(\tau,\bar\tau)(C^4+(F_5)^8+\cdots)
\end{equation}
\begin{equation}
\label{BFF}
S^{(3)}_{0;BFF}=\int d^{10}x\,f^{(0,0)}(\tau,\bar\tau)
(C^2\bar\psi\psi+(F_5)^7\bar\psi\psi\cdots)
\end{equation}
where $C$ is the Weyl tensor, $\psi$ is the gravitino, $F_5$ is the
self-dual five form and $f^{(0,0)}(\tau,\bar\tau)$ is a modular
function of the complex scalar fields $\tau$ and $\bar\tau$, which
reads as in (\ref{modfunc}, \ref{tau}). Note that the five form and
the metric are the only bosonic fields neutral under $U(1)$.  Now if
one starts from this restricted set of fields, considering terms only
linear in the fermions in the supersymmetry variations and setting
$\partial\tau=\partial\bar\tau=\lambda=0$ ($\lambda$ being the dilatino
of the theory), it is possible to show that the supersymmetry
variation of (\ref{B}) cancels exactly against the supersymmetry
variation of (\ref{BFF}) (neglecting fermions trilinear). Of course, 
setting the derivative of the scalar field $\tau$ to zero we are
effectively neglecting the variation of the modular form
$f^{(0,0)}(\tau,\bar\tau)$. Restricting our attention to these terms,
it is pretty straightforward to show that the obstruction to the
existence of the chiral measure, found in \cite{deHaro}, is
circumvented.  As one would expect, then, the final result for the eight
derivative action \cite{Rajaraman}, turns out to be exact.
This construction is, however, highly non-trivial and only few explicit
calculations were carried out \cite{Green,Green3}.  Recently, a
simplified, explicit expression for the eight derivative coupling
between the metric and five-form was found in \cite{Paulos}. In the
following we will make use of this general result, together with the
solutions (\ref{metunc}),(\ref{B4}). The full action reads:

\be\label{10dacn} 
I= \frac{1}{ 16\pi G_N} \int_{\calm_{10}} d^{10} x
\sqrt {-g} \ \bigg[ R_{10} - \tfrac12(\partial \phi)^2 - \frac{1}{4
  \cdot 5!}  (F_5)^2 +\cdots+ (\alpha')^3 \gamma(\phi)\mathcal{W} +
\cdots\bigg] 
\ee 
\be\label{defg} 
\gamma(\phi) = \frac{1}{16}
f^{(0,0)}(\tau,\bar{\tau}), \qquad G_N\propto \alpha'^4 
\ee
\be\label{modfunc} 
f^{(0,0)}(\tau,\bar \tau)= \sum_{(m,n)\neq(0,0)}
\frac{\tau_2^{3/2}}{|m+n \tau|^{3}}, 
\ee 

where
 \be\label{tau} \tau =
\tau_1 + i \tau_2 = C^{(0)}+ i e^{-\phi}.  
\ee

The above correction to the leading supergravity action is a complete
quantum (i.e. $\alpha'$ as well as string-loop) correction. The string
coupling $g_s \propto \exp{\phi_\infty}$.  The first term in the
expansion of $f^{(0,0)}(\tau,\bar \tau)$ appears only as a
supergravity ($\alpha'$) correction to the leading two derivative
Lagrangian , i.e. in finite $\alpha'$ and $N \rightarrow \infty$ limit
and thus, in $g_s \rightarrow 0 $ limit. For the following
computation, we will keep the entire quantum correction to the leading
supergravity action.

The higher derivative $\mathcal{W}$ contribution is explicitly given
by \cite{Paulos}:
\be\label{WT} \mathcal{W} \equiv
\frac{1}{86016}\sum_i n_i M_i \ee
\begin{center}
\begin{tabular}{|c|c|}
  \hline $n_i$  & $M_i$ \\ \hline
  -43008 & $ C_{a b c d} C_{a b e f} C_{c e g h} C_{d g f h} $ \\ \hline
  86016 & $  C_{a b c d} C_{a e c f} C_{b g e h} C_{d g f h}$ \\ \hline
  129024 & $ C_{a b c d} C_{a e f g} C_{b f h i} \mathcal T_{c d e g h i}$ \\ \hline
  30240 & $ C_{a b c d} C_{a b c e} \mathcal T_{d f g h i j} \mathcal T_{e f h g i j} $ \\ \hline
  7392 & $ C_{a b c d} C_{a b e f} \mathcal T_{c d g h i j} \mathcal T_{e f g h i j}$ \\ \hline
  -4032 & $ C_{a b c d} C_{a e c f} \mathcal T_{b e g h i j} \mathcal T_{d f g h i j}$ \\ \hline
  -4032 & $ C_{a b c d} C_{a e c f} \mathcal T_{b g h d i j} \mathcal T_{e g h f i j}$ \\ \hline
  -118272 & $ C_{a b c d} C_{a e f g} \mathcal T_{b c e h i j} \mathcal T_{d f h g i j}$ \\ \hline
  -26880 & $ C_{a b c d} C_{a e f g} \mathcal T_{b c e h i j} \mathcal T_{d h i f g j}$ \\ \hline
  112896 & $ C_{a b c d} C_{a e f g} \mathcal T_{b c f h i j} \mathcal T_{d e h g i j}$ \\ \hline
  -96768 & $ C_{a b c d} C_{a e f g} \mathcal T_{b c h e i j} \mathcal T_{d f h g i j}$ \\ \hline
  1344 & $ C_{a b c d} \mathcal T_{a b e f g h} \mathcal T_{c d e i j k} \mathcal T_{f g h i j k}$ \\ \hline
  -12096 & $ C_{a b c d} \mathcal T_{a b e f g h} \mathcal T_{c d f i j k} \mathcal T_{e g h i j k} $ \\ \hline
  -48384 & $ C_{a b c d} \mathcal T_{a b e f g h} \mathcal T_{c d f i j k} \mathcal T_{e g i h j k}$ \\ \hline
  24192 & $ C_{a b c d} \mathcal T_{a b e f g h} \mathcal T_{c e f i j k} \mathcal T_{d g h i j k} $ \\ \hline
  2386 & $ \ \mathcal T_{a b c d e f} \mathcal T_{a b c d g h} \mathcal T_{e g i j k l} \mathcal T_{f i j h k l}$ \\ \hline
  -3669 & $ \ \mathcal T_{a b c d e f} \mathcal T_{a b c d g h} \mathcal T_{e i j g k l} \mathcal T_{f i k h j l} $ \\ \hline
  -1296 & $ \ \mathcal T_{a b c d e f} \mathcal T_{a b c g h i} \mathcal T_{d e j g k l} \mathcal T_{f h k i j l} $ \\ \hline
  10368 & $ \ \mathcal T_{a b c d e f} \mathcal T_{a b c g h i} \mathcal T_{d g j e k l} \mathcal T_{f h k i j l} $ \\ \hline
  2688 & $ \ \mathcal T_{a b c d e f} \mathcal T_{a b d e g h} \mathcal T_{c g i j k l} \mathcal T_{f j k h i l}$ \\ \hline 
\end{tabular}\label{FinalRes}
\end{center}
The tensor $\mathcal T$ is defined by 
\be \mathcal T_{abcdef} =
P_{1050^+} \lb i \nabla_a F_{bcdef}+\frac18 F_{abcmn}F_{def}^{mn} \rb
.  \ee 
If we impose self-duality of the five-form, this reduces to
$$
\mathcal T_{a b c d e f}=i \nabla_{a} F_{b c d e f}+\frac 1{16}\left
  (F_{a b c m n}F_{d e f}^{\ \ \ m n}-3 F_{a b f m n}F_{d e c}^{\ \ \
    m n}\right), 
$$
where the RHS should be antisymmetrized in the triplets $[abc],[def]$ and
symmetrized for their interchange. Here, we also note that  
the higher derivative correction has been given in the Einstein
frame. We can also go to the string frame with proper  
transformation of the metric, but, obviously, the physical information 
of the system will not depend on the frame chosen.

\subsection{The fate of the flat direction}\label{fateofflat10d}

As we already explained, the dilaton parametrizes a flat direction at two derivative level, 
that can be lifted or not by the presence of supersymmetric higher derivative interactions. To verify its fate, we
first compute the entropy
function in presence of these higher derivative terms (\ref{WT}) and then focus on the attractor equations corresponding to the two
scalars, i.e. axion and dilaton.  Since the higher derivative term
$\mathcal{W}$ evaluated on the leading solution turns out to be
constant\footnote{ The value of $\mathcal{W}$ for the near-horizon geometry (\ref{nhparam}) considered is $14580$.}, the
axion-dilaton equations take the following form,
\be
\frac{\partial f^{(0,0)}(\tau_1,\tau_2)}{\partial
  \tau_1}\bigg|_{\tau_1=(\tau_1)_h, \tau_2=(\tau_2)_h}=0,  \quad
\frac{\partial f^{(0,0)}(\tau_1,\tau_2)}{\partial
  \tau_2}\bigg|_{\tau_1=(\tau_1)_h,\tau_2=(\tau_2)_h}=0,  
\ee
The axion equation takes the following form:
\be
\displaystyle\sum_{(m,n)\neq (0,0)}\frac{n (m+n \tau_1)}{|m+n \tau|^5}=0.
\ee
and it's easily solved by $\tau_1=0$. On the other hand, the dilaton 
equation of motion is given by (setting $\tau_1$ to zero):
\be
\displaystyle\sum_{(m,n)\neq (0,0)}\frac{\sqrt \tau_2(m^2- n^2 \tau_2^2)}
{|m^2+n^2\tau_2|^{5/2}}=0.
\ee
One solution of the above equation is $\tau_2=0 \implies \phi_h
\rightarrow \infty$, but this divergent behaviour destabilizes the near-horizon
geometry, so we won't take it into account. Another possible
solution is $\tau_2 =1$, for which
\be
\displaystyle\sum_{(m,n)\neq (0,0)}\frac{(m^2- n^2)}
{|m^2+n^2|^{5/2}}=0,
\ee
is identically satisfied. Therefore, the leading near-horizon value of
the dilaton is $\phi_h=0$, so that the flat
direction is lifted when we add higher derivative terms in
the action.

One important observation is that considering only the leading term in
the modular function, that is, only the leading higher derivative
correction ($\alpha'^3$) to supergravity, setting all loop correction
to zero, then the leading value of the dilaton is fixed to
infinity. Thus the thermodynamics of the system (temperature, entropy)
does not receive any correction due to this leading higher derivative
term, but the system is destabilized.

However, considering the full quantum correction, then there is a
possibility of a finite dilaton solution. This is a rather
interesting phenomenon, as the full quantum correction stabilizes the
system again. Not only that, it seems that supersymmetries, and not just extremality of a black hole solution is necessary to protect the flat directions
from being lifted. Once again, it looks as if the symmetries of the higher derivative interactions do not play any role to decide the fate of flat directions.

\subsection{Higher derivative correction to entropy}

For completeness, we compute higher derivative correction to
the entropy. We define the entropy function and the attractor equations as
before. We computed the full supersymmetric higher derivative term 
(\ref{WT}) for the near-horizon geometry and the expression is
presented in appendix \ref{dilatonsolution}. Solving the corrected attractor
equations we get the following corrections to the near-horizon
geometry:
\ben
v_1 &=& {\sqrt{-Q}\over 24}-\frac{144155}{384 Q}\hat{\gamma}, \quad v_2 =
{(2 \Theta/3)^{2/3} \over (-Q)^{5/6}} -\frac{25115 (\Theta
  (-Q))^{2/3}}{8 \ \sqrt[3]{2} \ 3^{2/3} Q^3}\hat{\gamma} ,\nn\\
z_1 &=&
{1\over 3 \sqrt{2}}-\frac{810\sqrt2}{(-Q)^{3/2}}\hat{\gamma}\ ,
\quad b = \frac{4\Theta}{3 Q}+ \frac{54115 \Theta }{9
   (-Q)^{5/2}}\hat{\gamma} , \quad w_1 =
 \frac{\sqrt{-Q}}{2}+\frac{21085}{32 Q}\hat{\gamma}
\een
where,
\be
\hat{\gamma}= \frac{\alpha'^3}{16} \sum_{(m,n)\neq (0,0)}
\frac{1}{(m^2 +n^2)^{3/2}}.
\ee
The entropy is given by:
\be
{\cal S} =\frac{\Theta}{192 \sqrt2 G_{10}} -\frac{405 \Theta
}{16 \sqrt{2} G_{10} (-Q)^{3/2}} \hat{\gamma}\ .
\ee

\section{Future Directions}\label{futuredirections}

We hope this work will pave the way for a number of possible applications
and extensions. Here we will mention some of them and hope to report
on them in future.

As of now, we are not aware of any asymptotically $AdS$ black hole solutions of
supergravity in ten dimensions that preserves some
supersymmetries. Thus, one interesting direction is to uplift the five
dimensional spinning $AdS$ solution, analyzed in the first part of this paper,
 to ten dimensions.  Knowing the
correct uplift of this class of supersymmetric five dimensional black
hole solutions, one can study the behavior of flat directions (if any) for the uplifted
solution and compare the results obtained with the ones presented in this paper.

In our work, we have seen that, for the five dimensional case, the
supersymmetry invariance of the higher derivative terms did not play
any role. In fact, without specifying the correct supersymmetric coefficients
of various higher derivative terms, the flat direction of the
leading solution remains flat. As we stressed before, the supersymmetries preserves by the leading
solution played an important role in the whole analysis. It would be
interesting to find a supersymmetric black hole solution in the higher
derivative theory as well although this would require an analysis of
the complete off-shell formulation of minimal gauged
supergravity in five dimensions. This analysis would certainly make use of the correct
values of various coefficients of the higher derivative terms and give
us the first supersymmetric asymptotic $AdS$ black hole solution away
from supergravity limit\footnote{Other theories, e.g. type IIB in 6 dimensions, admit flat directions and extremal black hole solutions \cite{Aste}; however for those theories the full, supersymmetric invariant, higher derivatives couplings are not known.}.

\vspace{1cm}
\bc
...............................
\ec

\vspace{3cm}

\noindent
{\bf Acknowledgement}
We would like to thank Bernard de Wit, Dileep Jatkar, Stefanos
Katmadas, Sameer Murthy, Bindusar Sahoo, Rajesh Gupta, Miguel Paulos,
Sera Cremonini and R. Loganayagam
for helpful discussion. We acknowledge our special thanks to Ashoke
Sen for his valuable comments on our work. We are very grateful to Jos
Vermaseren for his constant assistance and help in dealing with
computational and optimization issues in ``FORM'', the Symbolic Manipulation System he
developed and we used to compute higher-derivative terms in 10 D.
  I.L. would also like to thank Stijn van Tongeren, for his great
help at preliminary stages of this work.  The work of N.B. is supported
by an NWO Veni grant, the Netherlands. The work of I.L. is supported by
the ERC Advanced Grant no. 246974, \emph{"Supersymmetry: a window to
  non-perturbative physics"}.  S.D. would like to thank hospitality of
Nikhef during his visits. N.B. and S.D. would like to thank the
hospitality of String theory Discussion Meeting organized by ICTS. 
S.D. would also like to thank people of India for their generous support to
researches in basic sciences.

\appendix

\section{5D and 4D charges}\label{5d and 4d charges}

In this appendix, we provide the explicit expressions of the 5D and
the corresponding 4D charges for a $\frac{1}{4}$ BPS solution in
minimal gauged supergravity theory with supersymmetric higher derivative
terms. The results are given for the generic form of the solution in (\ref{nhgeoEF}) and for its supersymmetric
form in (\ref{nhgeoGR}). 
%At the end, we apply this formula for the special case we
%are interested in, i.e. the higher derivative corrected Gutowski-Reall
%spinning $AdS$ black holes and obtain its physical charges. 
First, the expressions for the 5D charges, which are completely general, are obtained using the equations
(\ref{electriccharge}), (\ref{NPE}), (\ref{NPAM}) and (\ref{angularmomenta}):

\begin{align}
  \label{q5}
  q_{5D}&= \frac{\pi}{G}\bigg[\frac{2 v_2 \sqrt{v_3}(e_1+e_0
    \varphi)}{v_1}- 8 \kappa \varphi(2 P+ B \varphi) + 32 c_2 v_2
  \sqrt{v_3}(e_1+e_0 \varphi)\bigg(\frac{(P+B \varphi)^2}{v_2{}^2
    v_1}-\frac{(e_1+e_0 \varphi)^2}{v_1{}^3}\bigg)
  \nn \\
  & \quad+ c_3 \frac{16 v_2 \sqrt{v_3}(e_1+e_0 \varphi)^3}{v_1{}^3} +
  4 c_4 \sqrt{v_3}\frac{v_2{}^2(e_1+e_0 \varphi)(4 v_1{}^2- 3 e_0^2
    v_3)
    + 2 B v_1{}^2 v_3 e_0( P+ B \varphi)}{v_1{}^3 v_2} \nn \\
  & \qquad \qquad + c_5 \frac{4 B v_3 (e_0^2v_2{}^2v_3+2
    v_1{}^2(v_2-B^2v_3))}{v_1{}^2v_2{}^2} \bigg]
\end{align}
Of course, adding suitable improvement terms to the current would have
led to the same exact result for the electric charge. Notice that
$\xi$ is just a constant, once we imposed the symmetric background
condition $\partial_\mu \xi=0$.  Similarly, the angular momenta takes the
following form:

\begin{align}
  \Theta_{5D} &= \frac{\pi}{G}\bigg[\frac{2 v_2 \sqrt{v_3}(e_0
    v_3+\varphi(e_1+e_0 \varphi))}{v_1} -8 \kappa \varphi^2 \frac{3 P+
    2 B \varphi}{3}+ 16 c_3 v_2 \sqrt{v_3}
  \varphi\frac{(e_1+e_0 \varphi)^3}{v_1{}^3} \nn \\
  & +2 c_1 e_0 v_3{}^{3/2}\frac{v_2{}^2(11 e_0{}^2 v_3 -12 v_1)-
    6 B^2 v_1{}^2 v_3}{v_1{}^3 v_2} \nn \\
  &+ 32 c_2 \sqrt{v_3} \varphi(e_1+e_0 \varphi)\frac{e_1 v_2{}^2(e_1+
    2 e_0 \varphi)- P v_1{}^2(P+ 2 B \varphi)
    + \varphi^2 (e_0{}^2 v_2{}^2- B^2 v_1{}^2)}{v_1{}^3 v_2}\nn \\
  &+4 c_4 \sqrt{v_3}\frac{2 v_1{}^2 v_3 (e_1+2 e_0 \varphi)(P+B
    \varphi)-v_2{}^2(e_1+e_0 \varphi)(3 e_0 v_3(e_1+2e_0\varphi)-4
    v_1\varphi)}{v_1{}^3 v_2} \nn \\
  &+ 4 c_5 v_3 \frac{P v_2{}^2(5 e_0{}^2 v_3- 4 v_1)+B(2e_1 e_0
    v_2{}^3 v_3+(9 e_0{}^2 v_2{}^2 v_3 - 4 v_1 v_2{}^2 + v_1{}^2(4 v_2
    - 3 B^2 v_3))\varphi)}{v_1{}^2 v_2{}^2} \bigg]
\end{align}

Analogously, plugging in the specific values of the parameters of the Gutowski-Reall solution, one obtains, for the corrected black hole charges:
\ref{chargescalc}
\ben
Q_5 & =& \frac{\sqrt{3} \Delta ^2 L^2 \left(\frac{3 L^2}{\Delta ^2 L^2-3}+\varphi ^2\right)+6 L
   \varphi -3 \sqrt{3} \varphi ^2}{12 G \left(\Delta ^2
     L^2-3\right)}-\frac{\gamma  \Delta ^2 L^4 }{48 G \left(\Delta ^2 L^2-3\right)^2
   \left(\Delta ^4 L^4-6 \Delta ^2 L^2-15\right)}\nn\\
&&  \bigg(\sqrt{3} c_1 (13 \Delta ^6 L^6-1179 \Delta ^4 L^4-7623
   \Delta ^2 L^2-135)-24 (2 c_5 \Delta ^6 L^6+65 \sqrt{3} c_4 \Delta ^4
   L^4\nn\\
&& -6 c_5 \Delta ^4 L^4+303 \sqrt{3} c_4 \Delta ^2 L^2+60 c_5 \Delta ^2 L^2+6
   \sqrt{3} c_2 (\Delta ^6 L^6+11 \Delta ^4 L^4+87 \Delta ^2
   L^2-99)\nn\\
&& +3
   \sqrt{3} c_3 (\Delta ^6 L^6+21 \Delta ^4 L^4+129 \Delta ^2 L^2+21)+450
   \sqrt{3} c_4-900 c_5)\bigg)
\een
\ben
\Theta_5 &=& \frac{9 \sqrt{3} \Delta ^2 L^4 \varphi +\sqrt{3} \varphi ^3 \left(\Delta ^2
   L^2-3\right)^2+9 L \varphi ^2 \left(\Delta ^2 L^2-3\right)-\frac{27 \Delta ^2
   L^5}{\Delta ^2 L^2-3}}{36 G \left(\Delta ^2 L^2-3\right)^2}\nn\\
&& +\frac{1}{48 G \left(\Delta ^2 L^2-3\right)^3 \left(\Delta ^4 L^4-6 \Delta ^2
   L^2-15\right)} (\gamma  \Delta ^2 L^4 (24 (6 \sqrt{3} c_2+3 \sqrt{3} c_3+2
   c_5)\nn\\
&& -13 \sqrt{3} c_1) \Delta ^8 L^8 \varphi +\Delta ^6 L^6 (c_1
   (635 L+1218 \sqrt{3} \varphi )+24 (6 c_2 (5 L+8 \sqrt{3}
   \varphi )+27 c_3 (L+2 \sqrt{3} \varphi )\nn\\
&&+36 c_4 L-4 \sqrt{3} c_5
   L+65 \sqrt{3} c_4 \varphi -12 c_5 \varphi ))-3 \Delta ^4 L^4
   ((795 c_1+816 c_2+552 c_3+1064 c_4\nn\\
&&+96 \sqrt{3} c_5) L-6 (227
   \sqrt{3} c_1+8 (54 \sqrt{3} c_2+33 \sqrt{3} c_3+18 \sqrt{3} c_4+13
   c_5)) \varphi )-9 \Delta ^2 L^2 ((449 c_1\nn\\
&&-816 c_2+120
   c_3+920 c_4-848 \sqrt{3} c_5) L+6 (421 \sqrt{3} c_1+960 \sqrt{3} c_2+488
   \sqrt{3} c_3+204 \sqrt{3} c_4\nn\\
&&+480 c_5) \varphi )-27 (175 c_1+8
   (26 c_2+13 c_3-20 c_4-30 \sqrt{3} c_5)) L-81 (5 \sqrt{3} c_1+8
   (-66 \sqrt{3} c_2\nn\\
&&+7 \sqrt{3} c_3+50 \sqrt{3} c_4-100 c_5)) \varphi
   ))
\een
Before we end this appendix, we summarize some of our
conventions for the 5D analysis. The metric is always mostly positive, with signature
$(-,+,+,+,+)$. The Levi-Civita tensor has the following form ,
\begin{equation}
 \epsilon ^{\mu \nu \rho\sigma \delta }= - \frac{\varepsilon^{\mu \nu \rho\sigma \delta }}{\sqrt{-g}}, \quad
 \epsilon _{\mu \nu \rho\sigma \delta }=  \sqrt{-g} \varepsilon_{\mu \nu \rho\sigma \delta },
\end{equation}
where, $\varepsilon^{\mu \nu \rho\sigma \delta }=\varepsilon_{\mu \nu
  \rho\sigma \delta }$ is just $\pm 1$ depending on the orientation of
the space-time. In a local Lorentz frame, $\epsilon _{01234}=1$. Also,
the definition of the Riemann tensor is
\begin{equation}
R^\mu\,_{\nu\rho\sigma}= 2\Gamma^\mu{}_{\nu[\sigma,\rho]}+ 2\Gamma^\lambda{}_{\nu[\sigma}\Gamma^\mu{}_{\rho]\lambda} .
\end{equation}

\section{Kaluza-Klein Reduction of 5D Lagrangian}\label{reduction}

In this appendix, we give the necessary details to perform the
Kaluza-Klein reduction of the five dimensional Lagrangian. The
reduction ansatz is given in (\ref{kkans}).\\
  In the tangent space, the
reduction of various components of the Riemann tensor reads:

\begin{align}
  \hat{R}_{cd}{}^{ab} &= \, R_{cd}{}^{ab} -\tfrac12 \phi \Big[
  F(B)_{c}{}^{[a} \,F(B)_d{}^{b]} + F(B)_{cd} F(B)^{ab}
  \Big]\,, \nonumber\\[.4ex]
 \hat{R}_{cd}{}^{a5} &=\,
  \phi^{\frac{1}{2}} \,D_{[c}(\omega)F(B)_{d]}{}^a +
  \Big[ \partial_{[c}\sqrt{\phi} \,F(B)_{d]}{}^a -
 \partial^a\sqrt{\phi}\, F(B)_{cd}\Big]  \,,  
\nonumber\\[.4ex]
\hat{R}_{c 5}{}^{ab}& =\, -\tfrac12\phi^{\frac{1}{2}} \, D_c(\omega)F(B)^{ab} -
  \Big[F(B)^{ab}\, \partial_{c}\sqrt{\phi} - F(B)_c{}^{[a}
  \,\partial^{b]}\sqrt{\phi} \Big]\,,   
\nonumber\\[.4ex]
\hat{R}_{c5}{}^{a5} &=\,-\phi^{-\frac{1}{2}}\, D_c(\omega)
  [\partial^a\sqrt{\phi}] + \tfrac14 \phi \,F(B)_{cb}F(B)^{ab}\,.
\end{align}
and gauge invariant two derivative Lagrangian takes the following form:
\begin{align}
S_0&=\,4 \pi \int d^4x \sqrt{-h} \sqrt{\Phi}
\Big(R+\frac{12}{L^2}-\frac{1}{4} F_{ij} F^{ij}
-\frac{1}{4}\left(\Phi +\sigma^2 \right)
(F_{kk})^{ij} (F_{kk})_{ij}  
\nonumber\\
& \quad -\frac{1}{2}\sigma F_{ij}(F_{kk})^{ij} + \frac{1}{2 \phi^2}
(\nabla \phi)^2- \frac{1}{\phi} \nabla^2 \phi - \frac{1}{2 \phi}
(\nabla \sigma)^2\Big) 
\end{align}

The higher derivative action consists of different four-derivative
terms . Here, we present the dimensional reduction of the gauge
invariant part:
\ben
(S_{\text{HD}})^{GI} &=& 4 \pi \int d^4x \sqrt{-h} \bigg[c_1 \, \bigg( \Phi^{1/2}  \
R^{{ijkl}} R_{{ijkl}} +\frac14 \Phi^{-7/2}  (\nabla_i
\phi \nabla^i \phi)^2 + \Phi^{-3/2}  \nabla^i\nabla^j\phi
\nabla_i\nabla_j \phi \nn\\
&&  - \frac32
\Phi^{3/2}  R_{ijkl} (F_{kk})^{ij}(F_{kk})^{kl} 
+\Phi^{1/2} 
(F_{kk})^i_{j}(F_{kk})^{jk} \nabla_k \nabla_i\phi - 2 \Phi^{-1/2} (F_{kk})^{jk}
(F_{kk})_{ij} \nabla_k \phi \nabla^i\phi \nn\\
&&+ \Phi^{3/2}  \nabla_i
(F_{kk})_{jk} \nabla^{i}(F_{kk})^{jk} 
 +3 \Phi^{1/2} \nabla_i
(F_{kk})_{jk} (F_{kk})^{jk} \nabla^i\phi -\Phi^{-5/2}  \nabla^i
\nabla^j e^{2 \Phi} \nabla_i\phi \nabla_j\phi \nn \\
&&+ \frac58 \Phi^{5/2} (F_{kk})^{ij}(F_{kk})_{jk}(F_{kk})^{kl}(F_{kk})_{li} -\frac32 \Phi^{-1/2}\nabla_i
\phi \nabla^i \phi (F_{kk})_{ij}(F_{kk})^{ij}  \bigg) \nn\\
&&+ c_2 \Phi^{1/2} \bigg(
F_{ij} F^{ij}  +\sigma^2
   (F_{kk})^{ij} (F_{kk})_{ij}  +2\sigma F_{ij}(F_{kk})^{ij} +\frac{2}{\Phi} (\nabla \sigma)^2\bigg)^2 
+  c_3\bigg(\Phi^{1/2} F^{ij}F_{jk}F^{kl}F_{li}\nn \\
&& + \Phi^{1/2} \sigma (F_{kk})^{ij}(F_{kk})_{jk}(F_{kk})^{kl}(F_{kk})_{li}
+ 4 \sigma  \Phi^{1/2} F_{ij} F^{jk} F_{kl} (F_{kk})^{li}
+  4 \sigma^2  \Phi^{1/2} F_{ij} F^{jk} (F_{kk})_{kl} (F_{kk})^{li} \nn \\
&&+  4 \sigma^3 \Phi^{1/2} F_{ij} (F_{kk})^{jk} (F_{kk})_{kl} (F_{kk})^{li}
 + 2 \sigma^2  \Phi^{1/2} F_{ij} (F_{kk})^{jk} F_{kl} (F_{kk})^{li}
+ 2  \Phi^{-3/2} \nabla_i (\sigma \nabla^i \sigma)^2 \nn\\
&&-4  \Phi^{-1/2} F_{ij} F^{jl} \nabla_l \sigma \nabla^i \sigma -4  \Phi^{-1/2} \sigma^2 F_{ij} (F_{kk})^{jl} 
\nabla_l \sigma \nabla^i \sigma -8  \Phi^{-1/2} \sigma (F_{kk})_{ij} (F_{kk})^{jl} \nabla_l \sigma \nabla^i \sigma\bigg) \nn\\
&&+ c_4 \bigg( \Phi^{1/2}R_{ijkl} F^{ij} F^{kl}  + \Phi^{1/2} \sigma^2  R_{ijkl} (F_{kk})^{ij} (F_{kk})^{kl} 
+ 2 \Phi^{1/2} \sigma  R_{ijkl} (F_{kk})^{ij} F^{kl} \nn\\
&&-\frac12 \Phi^{3/2} \big[\sigma^2 (F_{kk})^{ij}(F_{kk})_{jk}(F_{kk})^{kl}(F_{kk})_{li}
+ F^{ij}(F_{kk})_{jk}F^{kl}(F_{kk})_{li}+((F_{kk})^{ij} F^{ji})^2 \nn \\
&&+2 \sigma F^{ij}(F_{kk})_{jk}(F_{kk})^{kl}(F_{kk})_{li}
+2\sigma (F_{kk})^{ij} F_{ij} (F_{kk})_{kl}(F_{kk})^{kl}\big]
+ 4 \Phi^{1/2} \nabla_i (F_{kk})_{jk}F^{ij}\nabla^k \sigma \nn\\
&&+ 4 \Phi^{1/2}\sigma \nabla_i (F_{kk})_{jk}(F_{kk})^{ij}\nabla^k \sigma -\Phi^{1/2}\nabla^i \sigma (F_{kk})_{ij} (F_{kk})^{jk} 
\nabla_k  \sigma +2 \Phi^{-1/2}\big[\sigma (F_{kk})_{kl}(F_{kk})^{lk} \nn\\
&&+((F_{kk}))^{ij} F_{ji})\big]\nabla_i \phi \nabla^i \sigma - 2 \Phi^{-3/2}\nabla_i \sigma \nabla_j \sigma 
\nabla^j\nabla^i \phi+ \Phi^{-5/2}(\nabla_i \Phi \nabla^i \sigma)^2 
\bigg)\nn\\
&& 2 \Phi^{-1/2} \nabla^i \phi (F_{ij}+ \sigma (F_{kk})_{ij})(F_{kk})^{jk}\nabla_k \sigma
+\bigg(\frac{3 c1}{8} \Phi^{5/2} +\frac{3 c4}{4}\sigma^2 e^{3\Phi} \bigg)
((F_{kk})_{ij}(F_{kk})^{ij})^2 
\bigg]
\een

The full Lagrangian contains the above gauge invariant parts
as-well-as the reduction of the two gauge non-invariant Chern-Simons terms, that we
have already presented in the main text. To performed the entropy function analysis 
we have made use of the full Lagrangian.

\section{Redundancy of higher derivative corrections} \label{v1v2redun} 

As we already pointed out in the main text, the leading five dimensional 
solution (\ref{leadingsolutions5d}) has certain redundancies. Keeping the geometry of the full
solution fixed, we could impose an ansatz for the corrections to the
leading solution parameters: 
\ben
v_1 &=& \frac{1}{\Delta ^2+9L^{-2}} +\delta \tilde{V_1}, \qquad v_2 =
\frac{1}{\Delta ^2-3L^{-2}}+\delta \tilde{V_2},\qquad
v_3 = \frac{\Delta}{\Delta^2-3L^{-2}} +\delta \tilde{V}_3 ,\nn\\ 
e_0 &=& -\frac{3 }{L
  \Delta} \frac{\Delta^2-3L^{-2}}{\Delta ^2+9L^{-2}} +\delta \tilde{E_0},
\qquad 
e_5 = \frac{\sqrt{3} \Delta}{\Delta ^2+9L^{-2}}+\delta \tilde{E_5},\nn\\
A_{\chi}&=& \frac{\sqrt{3}\sin\theta}{L(\Delta^2-3L^{-2})} +\delta
\tilde{\Upsilon}, \qquad 
B= 1 +\delta \tilde{B_h}.
\een
Now, one can show that the leading Einstein's equations of
motion are satisfied up to order $\delta$, if the following
conditions are satisfied:
\ben
\tilde{V_1} &=&\frac{\left(\Delta ^2 L^2-3\right)^2 \left(2 \Delta ^2 L^4 \tilde{B}_h+2 \sqrt{3} L
   \tilde{\Upsilon } \left(\Delta ^2 L^2+1\right)+\tilde{V}_3 \left(\Delta ^2
   L^2-3\right)^2\right)}{\left(\Delta ^2 L^2+9\right)^2 \left(\Delta ^4 L^4-6 \Delta
   ^2 L^2-15\right)} \equiv f_1(\tilde{V_3},\tilde{B_h},\tilde{\Upsilon})\nonumber\\
\tilde{V_2}&=&\frac{2 \Delta ^2 L^4 \tilde{B}_h+2 \sqrt{3} L \tilde{\Upsilon } \left(\Delta ^2
   L^2+1\right)+\tilde{V}_3 \left(\Delta ^2 L^2-3\right)^2}{\Delta ^4 L^4-6 \Delta ^2
   L^2-15}\equiv
 f_2(\tilde{V_3},\tilde{B_h},\tilde{\Upsilon})\nonumber\\
\tilde{E_0} &=& \frac{1}{6 \Delta ^3 L^5 (\Delta ^2 L^2+9)^2 (\Delta
   ^4 L^4-6 \Delta ^2 L^2-15)}   ((\Delta ^2 L^2-3) (3 (16 \Delta ^4 L^6 (11 \Delta ^2
   L^2-9) \tilde{B}_h\nonumber\\
&&+\tilde{V}_3 (\Delta ^2 L^2-3)^2 (3 \Delta
   ^6 L^6+97 \Delta ^4 L^4-279 \Delta ^2 L^2-405))+2 \sqrt{3} \Delta ^2 L^3
   \tilde{\Upsilon } (3 \Delta ^8 L^8+8 \Delta ^6 L^6\nn\\
&&+54 \Delta ^4 L^4-288 \Delta
   ^2 L^2-81)))\equiv f_3(\tilde{V_3},\tilde{B_h},\tilde{\Upsilon})\nn\\
\tilde{E_5}&=& \frac{1}{2 \Delta  L^2 (\Delta ^2 L^2+9)^2 (\Delta ^4 L^4-6 \Delta ^2
   L^2-15)}(\sqrt{3} (\Delta ^6 L^6-5 \Delta ^4 L^4+147 \Delta ^2 L^2-135) (2
   \Delta ^2 L^4 \tilde{B}_h\nn\\
&& +\tilde{V}_3 \left(\Delta ^2 L^2-3\right)^2)+16
   \Delta ^2 L^3 \tilde{\Upsilon } (\Delta ^6 L^6+9 \Delta ^2 L^2-54)) \equiv
 f_4(\tilde{V_3},\tilde{B_h},\tilde{\Upsilon})
\een
The above solution is not supersymmetric and it is
specified in terms of four parameters: $\Delta, \tilde{V_3},
\tilde{\Upsilon}, \tilde{B_h}$. 

Now, when we add the higher derivative corrections to the leading solution
(\ref{corrections5d}), we could in principle add correction terms to
$v_3, B$ and $A_{\chi}$ also. Let the correction terms to these
quantities be $V_3, B_h$ and $\Upsilon$ respectively. We can plug
these corrected geometry to Einstein's equations but solve for only
four of them, because only four
equations are linearly independent.  We can choose to solve for $V_1,
V_2, E_0$ and $E_5$ in terms of other three parameters. The final
result is the following,
\ben
v_1 = v_1^{(0)} + \delta f_1(\tilde{V_3},\tilde{B_h},\tilde{\Upsilon})
+ \gamma f_1({V_3},{B_h},{\Upsilon}) + \gamma g_1 (c_i's)\nn\\
v_2 = v_2^{(0)} + \delta f_2(\tilde{V_3},\tilde{B_h},\tilde{\Upsilon})
+ \gamma f_2({V_3},{B_h},{\Upsilon}) + \gamma g_2 (c_i's)\nn\\
e_0 = e_0^{(0)} + \delta f_3(\tilde{V_3},\tilde{B_h},\tilde{\Upsilon})
+ \gamma f_3({V_3},{B_h},{\Upsilon}) + \gamma g_3 (c_i's)\nn\\
e_5 = e_5^{(0)} + \delta f_4(\tilde{V_3},\tilde{B_h},\tilde{\Upsilon})
+ \gamma f_4({V_3},{B_h},{\Upsilon}) + \gamma g_4 (c_i's)
\een
Now we can use the
redundancy in the leading solutions to remove corrections terms from
$V_3, B, \Upsilon$, i.e. we can choose $\delta$ to be $\gamma$ and
$\tilde V_3 = - V_3, \tilde B_h = -B_h$ and $\tilde{\Upsilon}= -
\Upsilon$. Then, the functions $g_i's$ read:

  \ben g_1&=& \frac{1}{4
  \left(\Delta ^2 L^2+9\right)^2 \left(\Delta ^4 L^4-6 \Delta ^2
    L^2-15\right)} \bigg(L^2 (c_1 (\Delta ^8 \left(-L^8\right)+269
\Delta ^6 L^6-525 \Delta ^4
L^4  \nonumber \\
&& +3861 \Delta ^2 L^2+1080)+24 (c_4 \Delta ^8 L^8+13 c_4 \Delta ^6
L^6+2
\sqrt{3} c_5 \Delta ^6 L^6-40 c_4 \Delta ^4 L^4\nonumber \\
&& -30 \sqrt{3} c_5 \Delta ^4 L^4-126 c_4 \Delta ^2 L^2+120 \sqrt{3}
c_5 \Delta ^2 L^2+6 c_2 (\Delta ^8 L^8-\Delta ^6
L^6-33 \Delta ^4 L^4\nonumber \\
&& +57 \Delta ^2 L^2-24)+3 c_3 \left(\Delta ^8 L^8+3 \Delta ^6 L^6-39
  \Delta ^4 L^4-9 \Delta ^2 L^2-24\right)+90 c_4))\bigg) 
\een 

\ben g_2
&=& \frac{1}{4 \left(\Delta ^6 L^6-9 \Delta ^4 L^4+3 \Delta ^2
    L^2+45\right)}(L^2 (c_1 (443 \Delta ^4 L^4-210 \Delta ^2
L^2+45)+24 (18 c_4
\Delta ^4 L^4\nonumber\\
&& +4 \sqrt{3} c_5 \Delta ^4 L^4+38 c_4 \Delta ^2 L^2-30 \sqrt{3} c_5
\Delta ^2 L^2+6 c_2 (7 \Delta ^4 L^4-10 \Delta ^2 L^2+3)\nonumber
\\
&&+3 c_3 \left(9 \Delta ^4 L^4+2 \Delta ^2 L^2+3\right)-15 c_4))) 
\een

\ben g_3 &=& -\frac{1}{3 \Delta L \left(\Delta ^2 L^2+9\right)^2
  \left(\Delta ^4 L^4-6 \Delta ^2 L^2-15\right)} ((\Delta ^2 L^2-3)
(c_1 (\Delta ^8 L^8-657 \Delta ^6
L^6\nonumber\\
&& -14238 \Delta ^4 L^4-4347 \Delta ^2 L^2+1215)+6 (-6 c_4 \Delta ^8
L^8+3 \sqrt{3} c_5 \Delta ^8 L^8-173 c_4 \Delta ^6 L^6\nonumber \\
&& +9 \sqrt{3} c_5 \Delta ^6 L^6-2067 c_4 \Delta ^4 L^4-519 \sqrt{3}
c_5 \Delta ^4 L^4-3987 c_4 \Delta ^2
L^2+1755 \sqrt{3} c_5 \Delta ^2 L^2\nonumber\\
&&-12 c_3 (28 \Delta ^6 L^6+219 \Delta ^4 L^4+24 \Delta ^2 L^2+27)+24
c_2 (\Delta ^8 L^8-25 \Delta ^6 L^6-168
\Delta ^4 L^4\nonumber\\
&& +219 \Delta ^2 L^2-27)+405 c_4)))
 \een 
\ben g_4&=&
-\frac{1}{\sqrt{3} \left(\Delta ^2 L^2+9\right)^2 \left(\Delta ^4
    L^4-6 \Delta ^2 L^2-15\right)} (\Delta L^2 (c_1 (4 \Delta ^8
L^8-135 \Delta ^6 L^6-1332 \Delta ^4
L^4\nonumber\\
&& -21465 \Delta ^2 L^2-810)+6 (c_4 \Delta ^8 L^8-23 c_4 \Delta ^6
L^6+3
\sqrt{3} c_5 \Delta ^6 L^6-399 c_4 \Delta ^4 L^4\nonumber\\
&& +57 \sqrt{3} c_5 \Delta ^4 L^4-3393 c_4 \Delta ^2 L^2-1071 \sqrt{3}
c_5 \Delta ^2 L^2-36 c_3 (19 \Delta ^4 L^4+136
\Delta ^2 L^2+3)\nonumber\\
&& +72 c_2 (\Delta ^6 L^6-18 \Delta ^4 L^4-115 \Delta ^2 L^2+132)-6750
c_4+2835 \sqrt{3} c_5))) \een

\section{Dilaton equation and its solution}
\label{dilatonsolution}

We want to study the complete profile of the dilaton, when the supersymmetric
higher derivative contribution to the action is considered. The dilaton equation of
motion is given by.
\be
\frac{d}{dr}\lb \sqrt{-g_{10}} \phi'(r) \rb + \frac{\gamma}{2} e^{-\frac32 \phi(r)}\sqrt{-g_{10}}\mathcal{W}(r)=0.
\ee
One can solve this equation in the extremal, near-horizon limit, i.e $q=2
r_0^2$ and $\mu = \frac{27 q^2}{4}$ ($L=1$). The leading
solution is constant, and we denote it by $\phi_h$. Corrections to the solution
can be found perturbatively in $\gamma$.  Since the higher derivative part is
already of the order $\gamma$, we need to evaluate it only on the
leading black hole geometry, obtaining:
\be
\mathcal{W}(r)= \frac{180 \mu^4}{(r^2+ Q)^8}
\ee
Now, we denote the correction term by
$\psi(r)$, so that the dilaton can be written as $\phi(r)= \phi_h +
\gamma \psi(r)$. Plugging this value in the dilaton equation, we get
a second order differential equation for $\psi(r)$, which is solved by:
\ben
\psi(r)&=& \frac{1}{1024} \bigg(\frac{32 \left(8 C_1+243 q^2 e^{-\frac{3 \phi _h}{2}}\right)}{9 q
   \left(q-2 r^2\right)}-\frac{32 \log \left(2 r^2-q\right) \left(8 C_1+4617 q^2
   e^{-\frac{3 \phi _h}{2}}\right)}{81 q^2}\nn\\
&& +\frac{\log \left(4 q+r^2\right) \left(512
   C_1+243 q^2 e^{-\frac{3 \phi _h}{2}}\right)}{162 q^2}-\frac{6561 q^5 e^{-\frac{3
   \phi _h}{2}}}{10 \left(q+r^2\right)^5}-\frac{6561 q^4 e^{-\frac{3 \phi _h}{2}}}{8
   \left(q+r^2\right)^4}\nn\\
&& -\frac{2187 q^3 e^{-\frac{3 \phi _h}{2}}}{2
   \left(q+r^2\right)^3}-\frac{5589 q^2 e^{-\frac{3 \phi _h}{2}}}{4
   \left(q+r^2\right)^2}-\frac{4617 q e^{-\frac{3 \phi _h}{2}}}{2
   \left(q+r^2\right)}+\frac{3645}{2} e^{-\frac{3 \phi _h}{2}} \log
   \left(q+r^2\right)\bigg)+C_2\nn\\
\een
This solution is singular at horizon ($r_0^2=\tfrac{q}2$, in the
extremal limit), specifically there are two terms which become singular 
at $r=r_0$, i.e. $\log(2r^2 -q)$ and $\frac{1}{2r^2 -q}$. We
can choose $C_1$ to set one of these two terms to zero, but to remove the 
other singular term we need to set $e^{-\frac32 \phi_h}=0$.  Therefore
the higher derivative term destabilizes the leading order solution (if
we consider the term appearing in front of ${\cal W}$ to be
$\exp(-3/2 \phi)$ only.). We found the same result from the entropy
function calculation performed in section \ref{fateofflat10d}.
Finally, for sake of completeness we present the expression for $\mathcal{W}$ in terms of the near-horizon parameters. 
The result has been obtained using FORM, and it reads:

\begin{align}
86016\,\mathcal{W}_{\text{NH}}&=
     -\frac{464}{9} v_1^{-8}v_2^{-6}w_1^2 z_1^4 q_5^4
      +\frac{1341256}{27}  v_1^{-8}  v_2^{-3} w_1^3  z_1^6 q_5^2
      +\frac{293461}{18}  v_1^{-8} w_1^4 z_1^8
      \nonumber
      \\
&	-\frac{3094952}{27}  v_1^{-7}  v_2^{-3} w_1^2 z_1^4 q_5^2
      -\frac{640136}{9}  v_1^{-7} w_1^3 z_1^6
      +\frac{44608}{9}  v_1^{-6}  v_2^{-9} w_1^{-1} z_1^2 b^2 q_5^4
      \nonumber
      \\
&      +\frac{327872}{9}  v_1^{-6}  v_2^{-6} z_1^3 b q_5^3
      -1848  v_1^{-6}  v_2^{-6} z_1^4 b^2 q_5^2
      +\frac{1896832}{27}  v_1^{-6}  v_2^{-3} w_1 z_1^2 q_5^2
      \nonumber
      \\
&     -\frac{1986880}{27} v_1^{-6}  v_2^{-3} w_1 z_1^4 q_5^2
      -\frac{3561656}{27}  v_1^{-6}  v_2^{-3} w_1 z_1^5 b q_5
      +\frac{355936}{3}  v_1^{-6} w_1^2 z_1^4
      \nonumber
      \\
&      +\frac{2990288}{9 } v_1^{-6} w_1^2 z_1^6
      +6720  v_1^{-5} v_2^{-6} w_1^{-1} z_1^2 b^2 q_5^2
      -\frac{1410752}{27}  v_1^{-5}  v_2^{-3} z_1^2 q_5^2
      \nonumber
      \\
&     +\frac{7592368}{27}  v_1^{-5}  v_2^{-3} z_1^3 b q_5
      -\frac{815360}{9}  v_1^{-5} w_1 z_1^2
      -\frac{2389184}{3}  v_1^{-5} w_1 z_1^4
      \nonumber
      \\
&     +4032  v_1^{-4}  v_2^{-12} w_1^{-4} b^4 q_5^4
      +\frac{122848}{9}  v_1^{-4} v_2^{-9} w_1^{-3} z_1 b^3 q_5^3
      +\frac{43192}{9}  v_1^{-4}  v_2^{-9} w_1^{-3} z_1^2 b^4 q_5^2
      \nonumber
      \\
&     -5376  v_1^{-4}  v_2^{-6} w_1^{-2} b^2 q_5^2
      +\frac{796736}{3}  v_1^{-4}  v_2^{-6} w_1^{-2} z_1^2 b^2 q_5^2
      +43008  v_1^{-4}  v_2^{-6} w_1^{-2} z_1^3 b^3 q_5
      \nonumber
      \\
&     -1764  v_1^{-4}  v_2^{-6} w_1^{-2} z_1^4 b^4
      -\frac{4046336}{27}  v_1^{-4}  v_2^{-3} w_1^{-1} z_1 b q_5
      +\frac{2308096}{27} v_1^{-4}  v_2^{-3} w_1^{-1} z_1^2 q_5^2
      \nonumber
      \\
&     +\frac{35321216}{27}  v_1^{-4}  v_2^{-3} w_1^{-1} z_1^3 b q_5
      -\frac{4528160}{27}  v_1^{-4}  v_2^{-3} w_1^{-1} z_1^4 b^2
      +\frac{250880}{9}  v_1^{-4}
      \nonumber
      \\
&     +\frac{1637888}{3}  v_1^{-4} z_1^2
      +\frac{3236800}{3}  v_1^{-4} z_1^4
      -13440  v_1^{-3}  v_2^{-9} w_1^{-4} b^4 q_5^2
      \nonumber
      \\
&     +10752  v_1^{-3}  v_2^{-6} w_1^{-3} b^2 q_5^2
      -43008  v_1^{-3}  v_2^{-6} w_1^{-3} z_1 b^3 q_5
      +6048  v_1^{-3}  v_2^{-6} w_1^{-3} z_1^2 b^4
      \nonumber
      \\
&     +\frac{10148992}{27}  v_1^{-3}  v_2^{-3} w_1^{-2} z_1 b q_5
      +\frac{9860032}{27}  v_1^{-3}  v_2^{-3} w_1^{-2} z_1^2 b^2
      -\frac{358400}{9} v_1^{-3} w_1^{-1}
      \nonumber
      \\
&     -\frac{1895936}{3}  v_1^{-3} w_1^{-1} z_1^2
      -2016  v_1^{-2}  v_2^{-12} w_1^{-6} b^6 q_5^2
      +\frac{875392}{9}  v_1^{-2}  v_2^{-9} w_1^{-5} b^4 q_5^2
      \nonumber
      \\
&     -\frac{204512}{9}  v_1^{-2} v_2^{-9} w_1^{-5} z_1 b^5 q_5
      +75264  v_1^{-2}  v_2^{-6} w_1^{-4} b^2 q_5^2
      -8064 v_1^{-2}  v_2^{-6} w_1^{-4} b^4
      \nonumber
      \\
&     +\frac{759296}{9}  v_1^{-2}  v_2^{-6} w_1^{-4} z_1 b^3 q_5
      +41664  v_1^{-2}  v_2^{-6} w_1^{-4} z_1^2 b^4
      -\frac{2551808}{27}  v_1^{-2}  v_2^{-3} w_1^{-3} b^2
      \nonumber
      \\
&     -\frac{5594624}{27}  v_1^{-2}  v_2^{-3} w_1^{-3} z_1 b q_5
      -\frac{16597504}{27}  v_1^{-2} v_2^{-3} w_1^{-3} z_1^2 b^2
      +\frac{501760}{3}  v_1^{-2} w_1^{-2}
      \nonumber
      \\
&     +\frac{11992064}{9}  v_1^{-2} w_1^{-2} z_1^2
      -5376  v_1^{-1}  v_2^{-6} w_1^{-5} b^4
      +\frac{555520}{27}  v_1^{-1}  v_2^{-3} w_1^{-4} b^2
      \nonumber
      \\
&     -\frac{1003520}{9}  v_1^{-1} w_1^{-3}
      +1512  v_2^{-12} w_1^{-8} b^8
      -\frac{260096}{9}  v_2^{-9} w_1^{-7} b^6
      \nonumber
      \\
&     -\frac{552704}{9}  v_2^{-6} w_1^{-6} b^4
      +\frac{17963008}{27 } v_2^{-3} w_1^{-5} b^2
      +\frac{2508800}{9} w_1^{-4}
\end{align}

\end{document}